\documentclass{elsart5p}

\usepackage{graphicx}
\usepackage{amssymb}   
\usepackage{textcomp}  
\usepackage{subfigure}

\let\oldequation\equation%
\let\endoldequation\endequation%
\renewenvironment{equation}{\vspace{4pt}\oldequation}{\endoldequation\vspace{4pt}}%

\let\olditemize\itemize%
\let\endolditemize\enditemize%
\renewenvironment{itemize}{\vspace{12pt}\olditemize}{\endolditemize\vspace{12pt}}%

\let\oldenumerate\enumerate%
\let\endoldenumerate\endenumerate%
\renewenvironment{enumerate}{\vspace{12pt}\oldenumerate}{\endoldenumerate\vspace{12pt}}%

\begin{document}

\begin{frontmatter}
\title{Spectral resolved Measurement of the Nitrogen Fluorescence Emissions in Air induced by Electrons}

\author[ad1]{T.~Waldenmaier\corauthref{cor1}}
\author[ad1,ad2]{J.~Bl\"umer},
\author[ad1]{H.~Klages}

\corauth[cor1]{Current address of corresponding author:\\
Bartol Research Institute, DPA, University of Delaware,\\
Newark, DE 19716 , U.S.A.\\
Email address: {\tt tilo@bartol.udel.edu} (Tilo Waldenmaier)}

\address[ad1]{Forschungszentrum Karlsruhe, Institut f\"ur Kernphysik, P.O.Box 3640, 76021 Karlsruhe, Germany}
\address[ad2]{Universit\"at Karlsruhe, Institut f\"ur Experimentelle Kernphysik, P.O.Box 6980, 76128 Karlsruhe, Germany}

\begin{abstract}
For the calorimetric determination of the primary energy of
extensive air showers, measured by fluorescence telescopes, a
precise knowledge of the conversion factor (fluorescence yield)
between the deposited energy in the atmosphere and the number of
emitted fluorescence photons is essential. The fluorescence yield
depends on the pressure and the temperature of the air as well as
on the water vapor concentration. Within the scope of this work
the fluorescence yield for the eight strongest nitrogen emission
bands between 300~nm and 400~nm has been measured using electrons
from a $\mathrm{^{90}Sr}$-source with energies between 250~keV and
2000~keV. Measurements have been performed in dry air, pure
nitrogen, and a nitrogen-oxygen mixture at pressures ranging from
2~hPa to 990~hPa. Furthermore the influence of water vapor has
been studied. A new approach for the parametrization of the
fluorescence yield was used to analyze the data, leading to a
consistent description of the fluorescence yield with a minimal
set of parameters. The resulting absolute accuracies for the
single nitrogen bands are in the order of 15~\%. In the
investigated energy range, the fluorescence yield proved to be
independent of the energy of the ionizing electrons.
\end{abstract}

\end{frontmatter}
%
%
%
\section{Introduction}
The measurement of air fluorescence is used by many modern
experiments (i.e. Pierre Auger Observatory~\cite{Abraham:2004},
HiRes~\cite{Springer:2005}) to detect extensive air showers (EAS),
induced by ultra-high energy cosmic rays. The secondary EAS
particles, predominantly electrons and positrons, deposit their
energy in the atmosphere by exciting or ionizing the air molecules
which afterwards may relax by emitting fluorescence photons. As
pointed out by Bunner~\cite{Bunner:1967} most of these emissions,
in the wavelength range between 300~nm and 400~nm, originate from
transitions of neutral or ionized nitrogen molecules. These faint
emissions can be measured by fluorescence telescopes, allowing the
observation of the longitudinal development of EAS through the
atmosphere and a calorimetric determination of the primary
cosmic-ray energy. The conversion factor between the deposited
energy in the air and the number of emitted fluorescence photons
is the so-called fluorescence yield $Y_{\lambda}(p,T)$ which
depends on the air pressure $p$ and temperature $T$ at the place
of emission, as well as on the wavelength $\lambda$ of the emitted
photons. If $Y_{\lambda}(p,T)$ does not depend on the energy of
the exciting particles the number of fluorescence photons,
observed by a telescope, is directly related to the deposited
energy according to
\begin{equation}\label{eq:ObservedPhotons}
    \frac{dN_{\gamma}}{dx} = \frac{dE_{dep}}{dx} \cdot
    \sum_{\lambda}  Y_{\lambda}(p,T)
    \cdot T_{atm}(\lambda,x) \cdot
    \varepsilon_{det}(\lambda,x)
\end{equation}
where $dx$ denotes an interval along the shower axis,
$T_{atm}(\lambda, x)$ is the atmospheric transmission factor and
$\varepsilon_{det}(\lambda, x)$ corresponds to the detection
efficiency of the telescope. The summation goes over all
wavelengths $\lambda$ passing the filter of the telescope,
typically ranging from 300~nm to 400~nm. The one dimensional
representation above only holds for large distances between the
EAS and the telescope where the lateral spread of the EAS can be
neglected. Integrating Eq.~(\ref{eq:ObservedPhotons}) over $x$
results in the total electromagnetic energy of the EAS and thus is
the most direct measure of the primary cosmic ray energy. The
precision of this method is, however, limited by the present
uncertainties of the fluorescence yield of about 15~\% to 30~\%
and the lack of knowledge about its energy dependence. In recent
years this gave rise to a number of new laboratory experiments,
i.e. Kakimoto et al.~\cite{Kakimoto:1995}, Nagano et
al.~\cite{Nagano:2003,Nagano:2004}, AIRFLY~\cite{Ave:2007},
FLASH~\cite{Huntemeyer:2003} or {\bf AirLight}, aiming at a
precise measurement of the fluorescence yield over a wide energy,
pressure and temperature range.

\begin{figure}[t]
\centering
\includegraphics[width=0.9\linewidth]{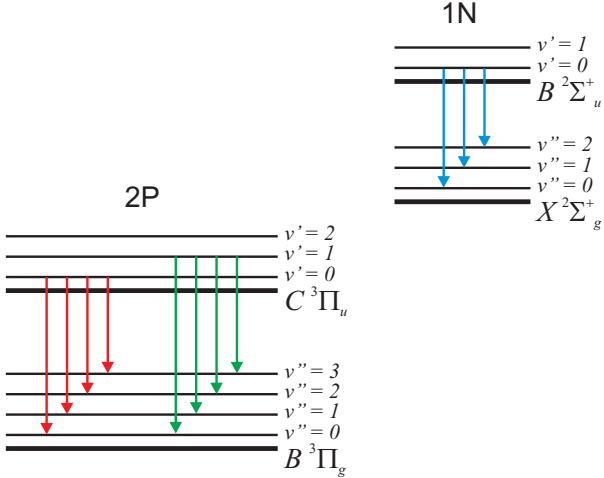}
\caption{Scheme of the energy states of the 2P and 1N
electronic-vibrational band systems of $\mathrm{N_{2}}$ and
$\mathrm{N_{2}^{+}}$. The corresponding spectral bands in
Fig.~\ref{fig:N2Spectrum} are drawn in the same
colors.}\label{fig:BandSystems}
\end{figure}

This paper reports on the data analysis and the results of the
{\bf AirLight} experiment at Forschungszentrum Karlsruhe in
Germany. Section~\ref{sec:NitrogenFluorescence} describes the
process of nitrogen fluorescence in air and introduces a new
approach to model the fluorescence yield in a consistent way and
with a minimal set of parameters. The experimental setup is
addressed in Section~\ref{sec:AirLight} and finally the data
analysis and the results are discussed in
Section~\ref{sec:Analysis}~and~\ref{sec:results}. This paper is
extracted from the Ph.D. thesis~\cite{Waldenmaier:2006} of the
corresponding author which can be referred for more details. Some
changes have been applied to correct minor errors and to improve the readability.
\section{Nitrogen Fluorescence in Air}\label{sec:NitrogenFluorescence}
\begin{figure}[t]
\centering
\includegraphics[width=\linewidth]{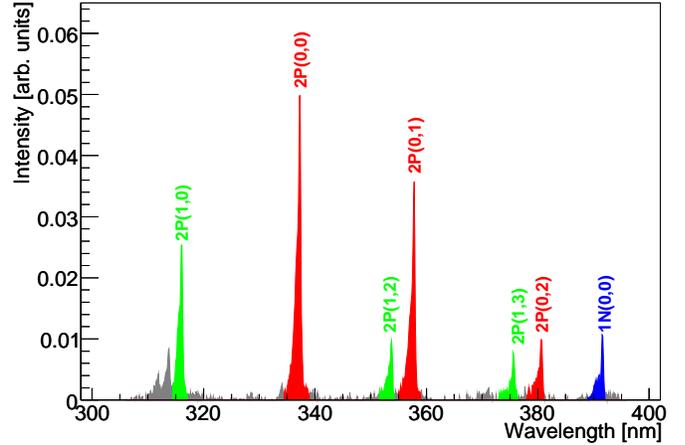}
\caption{Nitrogen fluorescence spectrum between 300~nm and 400~nm
in dry air at 1013~hPa measured by Ulrich et
al.~\cite{priv:Ulrich}.} \label{fig:N2Spectrum}
\end{figure}
Nearly all the air fluorescence emissions in the wavelength range
between 300~nm and 400~nm originate from neutral or ionized
nitrogen molecules~\cite{Bunner:1967,Pearse:1976}. The
fluorescence spectrum of molecular nitrogen is a band spectrum. In
contrast to atomic line spectra, molecular spectra consist of a
variety of broad bands. This band structure is caused by the
vibrational and rotational movements of the molecular nuclei which
modify the energy states of the electrons. For diatomic,
homo-nuclear molecules the energy of a molecular state can be
expressed as the sum of three contributions
\begin{equation}
    E = E_{el} + E_{vib} + E_{rot}\quad,
\end{equation}
where $E_{el}$ is the potential electron energy of a static
molecule and $E_{vib}$ and $E_{rot}$ are additional contributions
due to vibrations and rotations of the molecular nuclei. This
approach is known as the Born-Oppenheimer approximation
\cite{Bingel:1967,Haken:1992}. The energy scales of the three
contributions are very different and approximately behave like
\begin{equation}
    E_{el}:E_{vib}:E_{rot} =
    1:\sqrt{\frac{m}{M}}:\frac{m}{M}\quad,
\end{equation}
where $m$ is the electron mass and $M$ denotes the mass of the
molecular nuclei~\cite{Bingel:1967}. This causes an electronic
molecular state to split into several vibrational levels and every
vibrational level again has a rotational substructure, as is
illustrated in Fig.~\ref{fig:BandSystems}. Very often the
rotational substructure is not resolved, which leads to the
observation of vibrational bands with a sharp edge at one side and
a shading to the other side, as can be seen in
Fig.~\ref{fig:N2Spectrum}. Since the rotational substructure plays
no role for the application on EAS measurements it will be
disregarded in the following.

An "electronic band system" is established by all
electronic-vibrational transitions having the same initial and
final electronic states. The nitrogen fluorescence spectrum
between 300~nm and 400~nm nearly entirely consists of transitions
of the "second positive" (2P) band system\footnote{The notation
"first negative" and "second positive" system indicates the place
of appearance of the corresponding light emissions in gas
discharge tubes. "Negative" systems appear closer to the cathode
whereas "positive" systems are attracted by the anode.} of
$\mathrm{N_{2}}$ and the "first negative" (1N) band system of
$\mathrm{N_{2}^{+}}$~\cite{Bunner:1967,Pearse:1976}. The
spectroscopic notation for these transitions is given by
\begin{itemize}
    \item{2P-System:~~~~~$C^{3}\Pi_{u}(v') \rightarrow B^{3}\Pi_{g}(v'')$}\\
    \item{1N-System:~~~~~$B^{2}\Sigma_{u}^{+}(v') \rightarrow X^{2}\Sigma_{g}^{+}(v'')$}
\end{itemize}
where the quantum numbers $v'$ and $v''$ of the initial and final
vibrational levels are written within the parentheses behind the
symbols for the electron configuration of the corresponding
molecular state. Accordingly these transitions are denoted as
$\mathrm{2P}(v',v'')$ in case of the 2P-System and
$\mathrm{1N}(v',v'')$ for the 1N-System. In the following the
symbols $v'$ and $v''$ always denote initial and final
electronic-vibrational states of a certain electronic band system.
Any numerical values of $v'$ or $v''$ indicate the vibrational
level of the corresponding electronic state. Pure vibrational
transitions are not a subject of the following discussions.

\subsection{De-Excitation}\label{sec:de-excitation}
Assuming a number of excited nitrogen molecules in a certain
electronic-vibrational state $v'$ and neglecting any kind of
cascading into this state, the deactivation rate can be calculated
according to the general decay law
\begin{equation}\label{eq:decaylaw}
    \frac{dN_{v'}}{dt} = -\Bigr(\underbrace{\lambda_{rest} + \sum_{v''} A_{v',v''}}_{\lambda_{v'}}\Bigr)\cdot N_{v'}(t)\quad,
\end{equation}
where $N_{v'}(t)$ denotes the number of molecules in the state
$v'$ at a certain time $t$. The total decay constant
$\lambda_{v'}$, which equals the reciprocal mean lifetime
$\tau_{v'}$ of the state $v'$, is the sum of the decay constants
of all potential decay channels. The quantities $A_{v',v''}$
denote the decay constants or transition probabilities for all
radiative transitions $v' \rightarrow v''$ within the underlying
band system and are also known as Einstein coefficients. A
detailed list of Einstein coefficients of many nitrogen and oxygen
band systems can be found in the comprehensive work of Gilmore et
al.~\cite{Gilmore:1992}. In addition, there might be any kind of
other radiating or non-radiating transitions which are summarized
in the common decay constant $\lambda_{rest}$. From
Eq.~(\ref{eq:decaylaw}) two fundamental properties can be
immediately derived:
\begin{enumerate}\itemsep9pt
    \item The measurable mean lifetime $\tau_{v'}$ of an electronic-vibrational state $v'$ is the
    same for all transitions $v'~\rightarrow~v''$ originating from this state.
    \item The intensity ratios between spontaneous transitions of the same
    electronic-vibrational state $v'$ are constant and correspond to the
    ratios of their Einstein coefficients $A_{v',v''}$.
\end{enumerate}
The latter property causes the 2P and 1N band systems of the
nitrogen spectrum to be assembled from several sub-spectra for
each vibrational level $v'$, as is illustrated in
Fig.~\ref{fig:N2Spectrum}. In the wavelength range between 300~nm
and 400~nm the most intensive nitrogen transitions of the 2P
system originate from the vibrational levels $v'=0,1$, whereas the
1N system has just one notable transition 1N(0,0) at 391.4~nm. For
each of these sub-spectra the intensity ratios between their
vibrational bands are constant, but the absolute normalization of
each sub-spectrum in general changes differently with pressure and
temperature, as will be explained in the following.
\subsection{Radiationless Deactivations: Quenching}
An excited electronic-vibrational state $v'$ can also become
de-activated by radiationless processes such as rotational,
vibrational or translational energy transfer during collisions
with other molecules~\cite{Haken:1992}. These so-called quenching
processes strongly depend on the number
density~($\rightarrow$~Pressure) and the
velocity~($\rightarrow$~Temperature) of the colliding molecules.

Assume an excited nitrogen molecule in the state $v'$ hitting a
gas target where the target molecules are at rest. Under such
conditions a collisional deactivation constant
$\lambda_{c,x}^{v'}$ can be defined according to
\begin{equation}\label{eq:deactivationRate}
    \lambda_{c,x}^{v'} = n_{x}\cdot\sigma_{N_{2},x}^{v'}(v_{rel})\cdot v_{rel}\quad,
\end{equation}
where $n_{x}$ is the number density of the target molecules of
type $x$ and $\sigma_{N_{2},x}^{v'}(v_{rel})$ denotes the total
collisional cross-section for the deactivation of a nitrogen
molecule in the state $v'$ which was injected with a relative
velocity $v_{rel}$ with respect to the target molecules. In
reality, the projectile is part of the target gas and the
Maxwell-Boltzmann distribution applies for the velocities of the
target and projectile molecules. Therefore
Eq.~(\ref{eq:deactivationRate}) has to be averaged over all
relative velocities $v_{rel}$ between the colliding
molecules~\cite{Bunner:1967,Waldenmaier:2006}. The collisional
cross-sections $\sigma_{N_{2},x}^{v'}(v_{rel})$ in general depend
on the relative velocities $v_{rel}$ and thus on the temperature
of the gas. At present, no temperature dependence has been
reported and therefore $\sigma_{N_{2},x}^{v'}$ in the following is
assumed to be constant. In this approximation, averaging over the
molecular velocities in Eq.~(\ref{eq:deactivationRate}) leads to
the expression
\begin{equation}
    \lambda_{c,x}^{v'} = n_{x}\cdot\underbrace{\sigma_{N_{2},x}^{v'}\cdot\langle
    v_{N_{2}}\rangle\cdot\sqrt{\frac{m_{N_{2}} + m_{x}}{m_{x}}}}_{=:~Q_{x}^{v'}(T)}\quad,
\end{equation}
where $\langle v_{N_{2}}\rangle = \sqrt{8kT/\pi m_{N_{2}}}$ is the
mean velocity of the nitrogen molecules at temperature $T$
according to the Maxwell-Boltzmann distribution and $m_{x}$ is the
mass of the target molecules of type $x$. It is convenient to
introduce a quenching rate constant $Q_{x}^{v'}(T)$ for each gas
constituent $x$. Since $Q_{x}^{v'}(T)$ depends on $\langle
v_{N_{2}}\rangle$ it is not a real constant, but it is
proportional to $\sqrt{T}$ if no additional temperature dependence
from the collisional cross-sections $\sigma_{N_{2},x}^{v'}$ has to
be taken into account. All the quenching rate constants in this
paper are quoted for a temperature of
$293~K~(\sim20^{\circ}\mathrm{C})$. Quenching rate constants for
other temperatures can be computed according to
\begin{equation}\label{eq:QuenchRateTemp}
    Q_{x}^{v'}(T) = \sqrt{\frac{T}{293~K}}\cdot
    Q_{x}^{v'}(293~K)~~.
\end{equation}
If the gas is a mixture of different constituents $x$, the total
collisional deactivation constant $\lambda_{c}^{v'}$ is the sum of
the deactivation constants of all gas constituents:
\begin{eqnarray}\label{eq:collRateConstant}
    \lambda_{c}^{v'} = \sum_{x} \lambda_{c,x}^{v'} &=& \sum_{x} n_{x}\cdot
    Q_{x}^{v'}(T)\nonumber\\
    &=& \frac{p}{kT}\sum_{x} f_{x}\cdot Q_{x}^{v'}(T) =:
    C_{v'}\frac{p}{\sqrt{T}}\quad.
\end{eqnarray}
The last expression is proportional to $p/\sqrt{T}$ by a factor
$C_{v'}$ and was obtained by substituting the number densities
$n_{x}$ through the partial pressures $p_{x} = n_{x}kT$, as
follows from the ideal gas law. In a second step the partial
pressures $p_{x}$ have been expressed as $p_{x} = f_{x}\cdot p$,
where $f_{x}$ denotes the relative abundance of the different gas
constituents, e.g. for dry air $f_{N_{2}} = 0.78$, $f_{O_{2}} =
0.21$ and $f_{Ar} = 0.01$.

With this definition for the collisional deactivation constant
$\lambda_{c}^{v'}(p,T)$ the total decay constant or reciprocal
lifetime $1/\tau_{v'}$ of an electronic vibrational state $v'$ can
be expressed as
\begin{equation}\label{eq:totDecayConstant}
    \frac{1}{\tau_{v'}(p,T)} = \lambda_{c}^{v'}(p,T) + \underbrace{\lambda_{rest}^{v'} + \sum_{v''}
    A_{v',v''}}_{=:~1/\tau_{0}^{v'}}\quad,
\end{equation}
where $\lambda_{rest}^{v'}$ incorporates the decay constants for
all remaining deactivations or transitions with constant
transition probabilities. At zero pressure $\lambda_{c}^{v'}(p,T)$
vanishes and the total lifetime
$\left.\tau_{v'}(p,T)\right|_{p=0}$ becomes equal to the intrinsic
lifetime $\tau_{0}^{v'}$, which is the reciprocal sum of all
constant transition probabilities.

Instead of the quenching rate constants $Q_{x}^{v'}(T)$, it is
sometimes more convenient to introduce a reference pressure
$p'_{v'}(T)$ for a given gas mixture.  The reference pressure
$p'_{v'}(T)$ is defined as the pressure where the collisional
deactivation constant $\lambda_{c}^{v'}(p,T)$ equals the
reciprocal intrinsic lifetime $1/\tau_{0}^{v'}$. With this
definition Eq.~(\ref{eq:totDecayConstant}) transforms to the
simple form
\begin{equation}\label{eq:recLifetime}
    \frac{1}{\tau_{v'}(p,T)} = \frac{1}{\tau_{0}^{v'}}\cdot\left(1
    +
    \frac{p}{p'_{v'}(T)}\right)~~,~ p'_{v'}(T)=\frac{\sqrt{T}}{C_{v'}\tau_{0}^{v'}}~,
\end{equation}
which immediately reveals the linear pressure dependence of the
reciprocal lifetime $1/\tau_{v'}(p,T)$ for a constant temperature.
\subsection{Excitation: Intrinsic Fluorescence Yield}
The excitation of the relevant molecular nitrogen states in EAS
generally occurs via many different processes. The initial
electronic state $B^{2}\Sigma_{u}^{+}$ of the 1N system is usually
activated through the direct ionization of neutral nitrogen
molecules by high energy electrons
\begin{equation}
       \textrm{e} + \textrm{N}_{2}\left(X^{1}\Sigma_{g}^{+}\right) \longrightarrow
    {\textrm{N}^{+}_{2}}^{*}\left(B^{2}\Sigma_{u}^{+}\right) +
    \textrm{e}+\textrm{e} \quad,
\end{equation}
whereas the excitation of the $C^{3}\Pi_{u}$ state of the 2P
system mainly occurs via low energy secondary processes such as
the recombination of ionized nitrogen molecules
\begin{equation}
    \textrm{e} + \textrm{N}^{+}_{2}\left(X^{2}\Sigma_{g}^{+}\right) \longrightarrow
    {\textrm{N}_{2}}^{*}\left(C^{3}\Pi_{u}\right)\quad,
\end{equation}
or the direct excitation from the ground state
\begin{equation}
    \textrm{e}(\uparrow) + \textrm{N}_{2}\left(X^{1}\Sigma_{g}^{+}\right)\longrightarrow
    {\textrm{N}_{2}}^{*}\left(C^{3}\Pi_{u}\right) +
    \textrm{e}(\downarrow)\quad.
\end{equation}
The latter process, to the first order, is forbidden, since it
requires a change in the spin quantum number of the molecular
state. Nevertheless, this excitation becomes possible for low
energy electrons via electron exchange if spin-orbit-coupling is
taken into account~\cite{Bunner:1967,Haken:1992}.

In general, an electron releases its energy in a series of
different processes and only a small fraction of the energy, which
was deposited in a certain volume, is finally converted into
fluorescence photons. Thus if $\aleph_{v'}(E)$ is the number of
nitrogen molecules per deposited energy which have been excited
into the state $v'$, the fluorescence yield $Y_{v',v''}(E,p,T)$
for the radiative transition $v' \rightarrow v''$, in units of
photons per deposited energy, can be expressed as
\begin{eqnarray}\label{eq:flYield}
    Y_{v',v''}(E,p,T) &=& \aleph_{v'}(E)\cdot A_{v',v''}\cdot\tau_{v'}(p,T)\nonumber\\
    &~&~\nonumber\\
    &=& \underbrace{\aleph_{v'}(E)\cdot
    A_{v',v''}\cdot\tau_{0}^{v'}}_{=:~Y_{v',v''}^{0}(E)}\cdot\frac{\tau_{v'}(p,T)}{\tau_{0}^{v'}}\quad.
\end{eqnarray}
The term $A_{v',v''}\cdot\tau_{0}^{v'}$ in the final expression
determines the probability for an excited state $v'$ to relax via
the transition $v' \rightarrow v''$ if collisional quenching
effects are neglected. Therefore, the intrinsic fluorescence yield
$Y_{v',v''}^{0}(E)$ is defined as the number of emitted
fluorescence photons per deposited energy in the absence of
collisional quenching. The collisional quenching is finally taken
into account by the separate factor $\tau_{v'}(p,T)/\tau_{0}^{v'}$
which determines the fraction of excited states $v'$ which are not
going to be deactivated by collisional processes. This approach
naturally separates between excitation and de-excitation
processes. Potential energy dependencies of the effective
excitation process only affect the intrinsic fluorescence yield
$Y_{v',v''}^{0}(E)$, whereas pressure and temperature only act on
the quenching and thus on the lifetime $\tau_{v'}(p,T)$.

As pointed out in Section~\ref{sec:de-excitation}, the intensity
ratios between transitions emerging from the same
electronic-vibrational state $v'$ are always constant. In order to
apply this relation to Eq.~(\ref{eq:flYield}), a main-transition
$v' \rightarrow k$ needs to be specified for each initial
vibrational state $v'$. A main-transition is defined as the most
intensive transition of the corresponding vibrational sub-system,
but in principle the choice is completely arbitrary. In this work
the main-transitions for the investigated band systems have been
chosen to be 2P(0,0), 2P(1,0) and 1N(0,0). Now the intensity ratio
$R_{v',v''}$ of a transition $v' \rightarrow v''$ can be quoted
with respect to the main-transition as the ratio of the
corresponding Einstein-coefficients:
\begin{equation}
    R_{v',v''} = \frac{A_{v',v''}}{A_{v',k}}\quad,
\end{equation}
All main-transitions have $R_{v',k} \equiv 1$. With these
definitions Eq.~(\ref{eq:flYield}) transforms to the final
expression
\begin{equation}\label{eq:flYieldFinal}
    Y_{v',v''}(E,p,T) = Y_{v',k}^{0}(E)\cdot
    R_{v',v''}\cdot\frac{\tau_{v'}(p,T)}{\tau_{0}^{v'}}\quad,
\end{equation}
where $Y_{v',k}^{0}(E)$ denotes the intrinsic fluorescence yield
of the main-transition. All quantities in this equation now have a
clear physical meaning and are able to be measured experimentally.
Furthermore, all transitions emerging from the same initial state
$v'$ are treated consistently with a minimal set of parameters.

\begin{figure}[t]
\includegraphics[width=\linewidth]{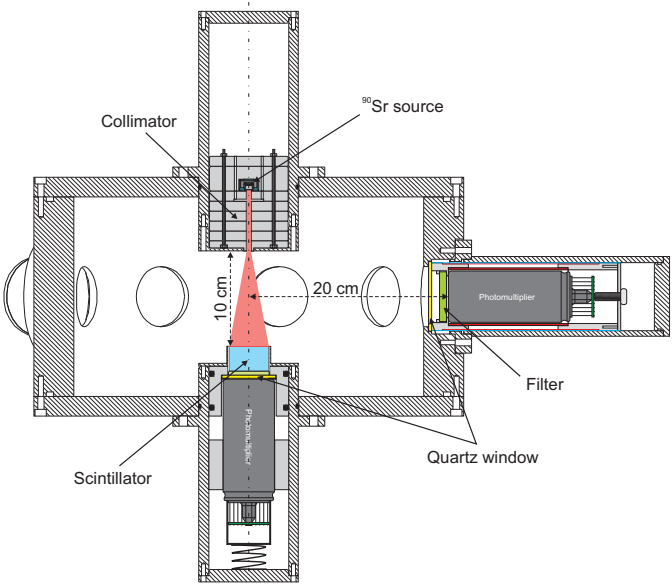}
\caption{Sketch of the AirLight chamber. After exiting the
collimator the electrons traverse 10~cm of gas before they are
stopped in the scintillator. Seven PMTs, equipped with different
filters, measure the fluorescence emissions perpendicularly to the
electron beam.} \label{fig:experiment}
\end{figure}

In previous papers the fluorescence yield is often defined as the
number of emitted fluorescence photons per meter track length of
the ionizing particle. This definition has several disadvantages
as will be explained in the following. First of all it requires
all secondary electrons to be tracked as well in order to obtain
correct results. Since the field of view of all laboratory
experiments is limited to a relatively small region around the
electron beam, secondary electrons can easily escape and thus do
not contribute to the fluorescence emissions in this volume. It is
even possible that electrons are scattered back from the chamber
walls into the field of view and induce additional fluorescence
photons. All these effects are very difficult to treat if the
fluorescence yield is defined as photons per meter. Instead it is
relatively simple to retrieve the deposited energy in a certain
volume by Monte Carlo simulations (see
Section~\ref{sec:EnergyDeposit}) and to determine the number of
fluorescence photons per deposited energy, which accounts for all
effects mentioned above. Another drawback of the definition in
photons per meter is that the fluorescence yield depends on the
underlying energy loss model as well as on the energy and the
pressure, which makes it rather difficult to compare values of
different authors. All these problems diminish if the fluorescence
yield is defined as photons per deposited energy, which seems also
to be more natural, since the deposited energy is the maximum
energy which may dissipate into fluorescence photons. This
definition is much more convenient, particularly for EAS
measurements, since it allows the number of observed photons to be
directly converted into an energy deposit profile of the EAS by
means of Eq.~\ref{eq:ObservedPhotons}. Therefore, the fluorescence
yield in this paper is always defined as photons per deposited
energy. The "old" definition in terms of photons per meter can be
obtained by multiplying the fluorescence yield with an appropriate
energy loss function $\frac{dE}{dx}$.
\section{The AirLight Experiment}\label{sec:AirLight}
\begin{figure}[t]
\includegraphics[width=\linewidth]{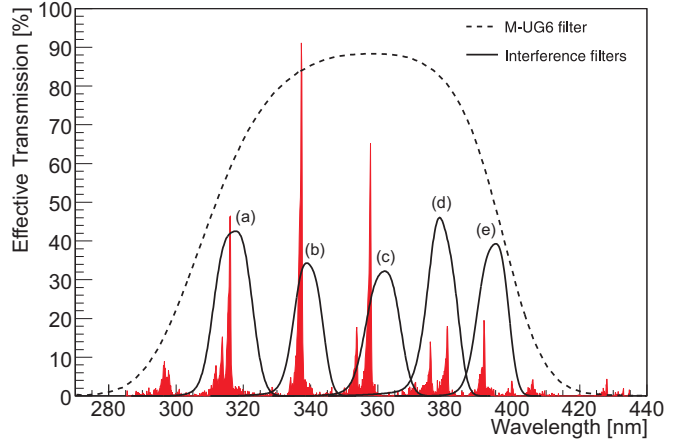}
\caption{Effective transmission curves of the interference filters
(solid lines) and the broad band M-UG6 filter (dashed line) overlaid with a nitrogen fluorescence spectrum measured by Ulrich et
al.~\cite{priv:Ulrich}.}
\label{fig:filters}
\end{figure}
The AirLight~experiment~\cite{Waldenmaier:2006} at
Forschungszentrum Karlsruhe was designed to measure all the
different quantities in Eq.~(\ref{eq:flYieldFinal}). The
experimental setup is similar to the experiments done by Kakimoto
and Nagano et al.~\cite{Kakimoto:1995,Nagano:2003,Nagano:2004}. As
is shown in Fig.~\ref{fig:experiment} it consists of a cylindrical
aluminum chamber in which electrons are injected along the chamber
axis. The electrons are emitted from a $\mathrm{^{90}Sr}$-source
situated at the top of the chamber. The source has an activity of
37~MBq with an end point energy of 2.3~MeV. The electrons are
collimated by massive lead rings with an inner diameter of 5~mm, before they enter the inner
volume of the chamber. After traversing 10~cm of gas (dry air, pure
nitrogen, or a nitrogen-oxygen mixture) they are finally stopped in a plastic scintillator of 2.5~cm height and 4~cm diameter~\cite{Waldenmaier:2006}. The plastic scintillator
measures the energy of the electrons with an energy resolution of
about 10~\% at 1~MeV. The electron rate at the scintillator varies
between 10~kHz and 20~kHz, depending on the pressure in the
chamber which can be adjusted between 2~hPa and 1000~hPa.

Due to the relatively low electron rate the intensity of the
electron-induced fluorescence light is not sufficient to allow the
use of a spectrometer or monochromator to make a spectral resolved
measurement. Therefore, seven $2^{\shortparallel}$ Photonis
photomultipliers\footnote{Photonis PMT types: XP2262,
XP2268}~(PMT) equipped with different filters are mounted
perpendicularly, at a distance of 20~cm, to the chamber axis to
measure the fluorescence photons in several wavelength ranges.
The active area of each of the photocathodes is defined by a 4~cm aperture. As
shown in Fig.~\ref{fig:filters}, the set of filters consists of
one broad band M-UG6 absorption filter ranging from 300~nm to
410~nm, identical to those used in the fluorescence telescopes of
the Pierre Auger Observatory~\cite{Abraham:2004}, and six narrow
band interference filters matched to the most prominent nitrogen
bands. Since interference filters are known to change their
transmission characteristics with the incident angle of the light,
the effective transmission functions plotted in
Fig.~\ref{fig:filters} were obtained by averaging over the angle
distribution of the fluorescence photons
\cite{Klepser:2004,Waldenmaier:2006} which was determined by means
of Monte Carlo simulations~\cite{Waldenmaier:2006}. The pressure dependence of the angular distribution, due to the multiple scattering of the electrons, turned out to be negligible.

\begin{figure}[t]
\includegraphics[width=\linewidth]{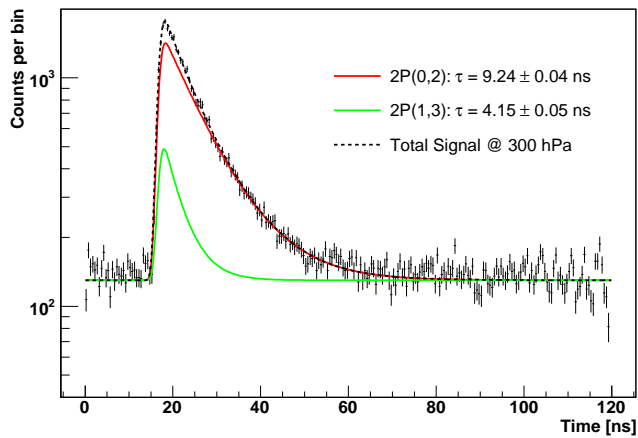}
\caption{Exponential time distribution between the electron
signals in the scintillator and the photon signals in channel~3.
The equally distributed offset is due to accidental coincidences
(thermal noise) in the PMT. The total fluorescence signal in this
channel is a superposition of the 2P(0,2) and the 2P(1,3)
transitions.} \label{fig:timeSpec}
\end{figure}

\begin{figure}[t]
\includegraphics[width=\linewidth]{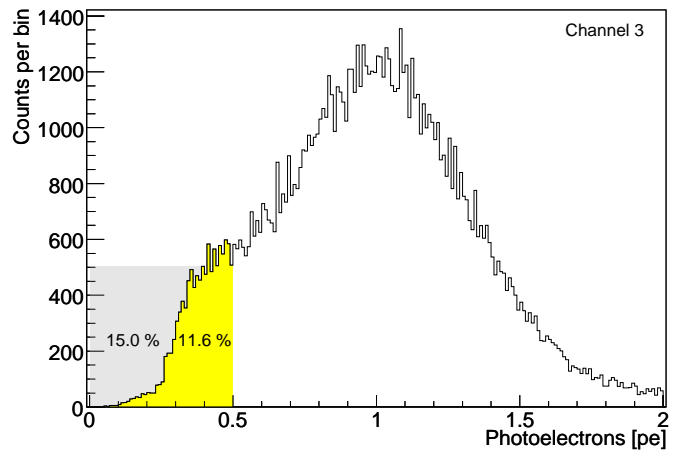}
\caption{Single photoelectron distribution of channel~3. Only
signals between 0.5~pe and 2~pe are considered in the data
analysis. Roughly 11.6~\% of the signals are between the
discriminator threshold and 0.5~pe (yellow region). The gray
region is a crude estimate of 15~\% for the signals below the
discriminator threshold.} \label{fig:peSpec}
\end{figure}

The experiment measures coincidences between individual electron
signals in the scintillator and single photon signals in any of
the PMTs. The coincidence condition is fulfilled if a photon
signal appears within an interval of 120~ns after an electron
signal was detected. The distribution of time differences~(see
Fig.~\ref{fig:timeSpec}) between electron and photon signals, as
well as the single photoelectron distributions~(see
Fig.~\ref{fig:peSpec}), are sampled for each filter channel by a
CAEN V488 TDC\footnote{TDC: Time to Digital Converter} and a
Lecroy 1182 ADC\footnote{ADC: Analog to Digital Converter}. One
ADC channel is sampling the energy distributions of the electrons.
After every tenth coincidence, the coincidence condition is
disabled for the next electron to measure also the unbiased (free)
electron energy spectrum. Finally the absolute numbers of free and
coincident photon and electron triggers are counted by a CAEN V260
scaler module with a guaranteed input frequency of 100~MHz. The
typical free rates of the photon detectors at
$\mathrm{20~^{\circ}C}$ ranged from 400~Hz to 600~Hz.

By means of the TDC time spectra, fluorescence signals can be
clearly discriminated from accidental coincidences. As is shown in
Fig.~\ref{fig:timeSpec} the time distribution of the fluorescence
photons follows an exponential decay law as is expected from
Eq.~(\ref{eq:decaylaw}). Contrary to this, accidental coincidences
show a flat distribution, since the timing of the electron and
photon signals is completely uncorrelated. The time resolution
$\sigma_{t}$ of each filter channel is of the order of 0.7~ns.
Therefore, the time distribution of the fluorescence signals can
be described by a convolution of an exponential- and a Gaussian
function with $\sigma=\sigma_{t}$ which is given by the expression
\begin{equation}\label{eq:TimeDistribution}
    \frac{dN}{dt} = \frac{1}{2}\frac{N}{\tau} \cdot e^{-\frac{t-t_{0}}{\tau}}
    \cdot e^{\frac{\sigma_{t}^{2}}{2\tau^2}}
    \cdot\mathrm{erfc}\left(\frac{t_{0} - t +
    \frac{\sigma_{t}^{2}}{\tau}}{\sqrt{2}\sigma_{t}}\right)\quad.
\end{equation}
The time integral of this function corresponds to the total number
of fluorescence photons $N$, whereas $\tau$ denotes the lifetime
of the emitting nitrogen state. As can be seen in
Fig.~\ref{fig:timeSpec} the time origin $t_{0}$ is in the order of
16~ns and is roughly the same in all filter channels. In general
there are several nitrogen bands with different lifetimes $\tau$
overlapping in one filter channel. Therefore, the measured time
distribution results from a superposition of several distributions
given by Eq.~(\ref{eq:TimeDistribution}) with different lifetimes
and intensities but with the same values for the time origin and
time resolution of the relevant filter channel.

The number of detected photons $N_{det}^{i}$ in a filter channel
$i$ is linked to the real number of emitted photons $N$ through
the relation
\begin{equation}\label{eq:calibration}
N_{det}^{i} =
N\cdot\underbrace{\varepsilon_{cut}^{i}\cdot\varepsilon_{col}^{i}\cdot f_{i}}_{f_{cal}^{i}}\cdot\varepsilon_{\Omega}\underbrace{\int_{\lambda}\varepsilon_{QE}^{0}(\lambda)\cdot
T(\lambda)\cdot\frac{d\hat{N}}{d\lambda}~d\lambda}_{\varepsilon_{s}^{i}}
\end{equation}
The integral in this expression corresponds to the spectral
efficiency $\varepsilon_{s}^{i}$, which contains the nominal
quantum efficiency function $\varepsilon_{QE}^{0}(\lambda)$ of the
PMT, as quoted by the manufacturer, the effective transmission
function $T(\lambda)$ of the filter and the normalized spectral
distribution $\frac{d\hat{N}}{d\lambda}$ of the emitted photons. If the spectral distribution is known, the integral $\varepsilon_{s}^{i}$ can be easily computed, since the unknown normalization factor $f_{i}$ of the individual quantum
efficiency function $\varepsilon_{QE}^{i}(\lambda) = f_{i}\cdot
\varepsilon_{QE}^{0}(\lambda)$ has been moved out of the integral.
The PMT acceptance is given by the geometrical factor
$\varepsilon_{\Omega}$ which is the same for all filter channels,
since the whole setup is completely symmetric against the chamber
axis. For fluorescence measurements $\varepsilon_{\Omega}$ has
been determined by means of Monte Carlo simulations to a value of
$(0.2586 \pm 0.0003)~\%$ which doesn't depend on the pressure. The collection efficiency $\varepsilon_{col}^{i}$ of the PMTs is assumed to be constant over the relatively small filter range. Finally we must account for the cut efficiency
$\varepsilon_{cut}^{i}$, since only photon signals between 0.5~pe
and 2.0~pe\footnote{[pe]: photoelectrons} are used in the data
analysis as illustrated in Fig.~\ref{fig:peSpec}. The
normalization factor $f_{i}$, the collection efficiency $\varepsilon_{col}^{i}$ and the cut efficiency are the only unknown quantities in Eq.~(\ref{eq:calibration}). For convenience they have been merged into the final calibration
constant $f_{cal}^{i}$.

\subsection{Absolute Calibration}

\begin{table*}[t]
\caption{Overview of the different filter channels and the
corresponding nitrogen bands. The detection efficiencies are
determined according to
$\varepsilon_{det}~=~f_{cal}\cdot\varepsilon_{\Omega}\cdot\varepsilon_{s}$,
using a common acceptance value of
$\varepsilon_{\Omega}~=~0.2586~\%$. The last column of the table
indicates the relative systematic errors
$\Delta\varepsilon_{det}$. }\label{tab:channels}
  \flushleft
  \begin{tabular}{cccccccc}
    \hline
    Channel & ~~~~~~~Filter~~~~~~ &  ~~~~~~~$f_{cal}$~~~~~~ & ~~~~~~Band~~~~~~ & ~~~~~~~$\lambda$~[nm]~~~~~~~ & ~~~~~~~~~~~$\varepsilon_{s}$~~~~~~~~~~~ & ~~~~~~~~~~$\varepsilon_{det}$~[\textperthousand]~~~~~~~~& ~~~~~~~~~$\Delta\varepsilon_{det}$~[\%]~~~~~~~\\
    \hline
    0 & M-UG6 &    $0.832 \pm 0.109$ & 2P(1,0) & 315.9 & $0.077 \pm 0.006$ & $0.166 \pm 0.025$ & $14.93$\\
    ~ & ~ & ~                        & 2P(0,0) & 337.1 & $0.158 \pm 0.008$ & $0.340 \pm 0.048$ & $14.09$\\
    ~ & ~ & ~                        & 2P(1,2) & 353.7 & $0.192 \pm 0.009$ & $0.413 \pm 0.057$ & $13.86$\\
    ~ & ~ & ~                        & 2P(0,1) & 357.7 & $0.196 \pm 0.009$ & $0.422 \pm 0.058$ & $13.84$\\
    ~ & ~ & ~                        & 2P(1,3) & 375.5 & $0.200 \pm 0.008$ & $0.430 \pm 0.059$ & $13.76$\\
    ~ & ~ & ~                        & 2P(0,2) & 380.5 & $0.194 \pm 0.008$ & $0.418 \pm 0.057$ & $13.74$\\
    ~ & ~ & ~                        & 1N(0,0) & 391.4 & $0.152 \pm 0.006$ & $0.327 \pm 0.045$ & $13.68$\\
    ~ & ~ & ~                        & 2P(1,4) & 399.8 & $0.085 \pm 0.003$ & $0.183 \pm 0.025$ & $13.68$\\
    \\
    1 & (a)   &    $0.714 \pm 0.094$ & 2P(1,0) & 315.9 & $0.094 \pm 0.004$ & $0.173 \pm 0.024$ & $13.99$\\
    \\
    2 & (c)   &    $0.825 \pm 0.108$ & 2P(1,2) & 353.7 & $0.014 \pm 0.001$ & $0.029 \pm 0.004$ & $15.39$\\
    ~ & ~ & ~                        & 2P(0,1) & 357.7 & $0.053 \pm 0.003$ & $0.112 \pm 0.016$ & $14.03$\\
    \\
    3$^{*}$ & (d)& $0.737 \pm 0.095$ & 2P(1,3) & 375.5 & $0.078 \pm 0.004$ & $0.148 \pm 0.020$ & $13.61$\\
    ~ & ~ & ~                        & 2P(0,2) & 380.5 & $0.106 \pm 0.005$ & $0.202 \pm 0.027$ & $13.53$\\
    \\
    4 & (b)   &    $0.854 \pm 0.114$ & 2P(0,0) & 337.1 & $0.060 \pm 0.003$ & $0.133 \pm 0.019$ & $14.38$\\
    \\
    5 & (e)   &    $0.788 \pm 0.104$ & 1N(0,0) & 391.4 & $0.081 \pm 0.004$ & $0.165 \pm 0.023$ & $13.84$\\
    ~ & ~ & ~                        & 2P(1,4) & 399.8 & $0.042 \pm 0.002$ & $0.085 \pm 0.012$ & $14.08$\\
    \hline
  \end{tabular}\\
  $^{*}$~Reference channel.\\
\end{table*}
In principle the calibration constants $f_{cal}^{i}$ can be
individually determined by a calibration measurement where a well
calibrated light source is placed in the center of the chamber. In
the present work it was not possible to perform such an end-to-end
calibration since an appropriate light source was not yet
available. Nevertheless a relative calibration has been performed
where an uncalibrated but stable and axial symmetric light source
with a well known spectrum was used instead. Furthermore the filters in front of the PMTs
have been removed, resulting in $T(\lambda) \equiv 1$ for each
channel. This way all PMTs are illuminated with an equal but
unknown amount of photons and the acceptance
$\varepsilon_{\Omega}$ is the same for all channels due to
symmetry reasons. By means of Eq.~(\ref{eq:calibration}) the
calibration constants $f_{cal}^{i}$ have been relatively
determined according to
\begin{equation}\label{eq:relCalibration}
    \frac{N_{det}^{i}}{N_{det}^{3}} = \frac{f_{cal}^{i}
    \cdot \varepsilon_{s}^{i}}{f_{cal}^{3}
    \cdot \varepsilon_{s}^{3}} \quad \Longleftrightarrow \quad f_{cal}^{i}
    = \frac{\varepsilon_{s}^{3}}{\varepsilon_{s}^{i}}\cdot\frac{N_{det}^{i}}{N_{det}^{3}}\cdot
    f_{cal}^{3}~~,
\end{equation}
where channel~3 was chosen for technical reasons to be the
reference channel. In case of equal PMT types the spectral
efficiencies $\varepsilon_{s}^{i}$ cancel and the calibration
constant is just the fraction of the detected photons with respect
to the reference channel times the reference constant. In order to evaluate absolute values for the calibration constants, the reference constant $f_{cal}^{3}$ was estimated
under the following conditions:

The shape of the single photoelectron distribution of the reference channel, shown in
Fig.~\ref{fig:peSpec}, as well as the discriminator threshold are
assumed to be stable over the whole period of measurements. Stable
photoelectron distributions are also required for all the other
channels. Furthermore, the photoelectron distributions of
accidental coincidences, which are mainly caused by thermal noise
of the PMTs, must be roughly the same as for real photon signals.
These requirements have been proven to be satisfactorily fulfilled
over the whole data-taking period~\cite{Waldenmaier:2006}.

The area between 0.5~pe and 2.0~pe in the single photoelectron
distribution shown in Fig.~\ref{fig:peSpec} corresponds to 100~\%
of the detected photons which have been considered in the
data-analysis. Signals above 2.0~pe are mainly due to noise or
background and can be neglected. However, roughly 12~\% of the
signals are situated between the discriminator threshold and the
quality cut at 0.5~pe (yellow region) and have to be taken into
account. The number of signals between the pedestal at 0~pe and
the discriminator threshold (gray region) is unknown but has been
conservatively estimated to be less than 15~\% of the detected signals.
Therefore, half of this value (7.5~\%) is added to the number of
detected signals and the other 7.5~\% are taken as the systematic
uncertainty of this estimation. Thus, if $N_{cut}$ denotes the
number of signals ranging from 0.5~pe to 2.0~pe, the real number
of signals $N_{real}$ can be evaluated by the expression
\begin{equation}
    N_{real} = (100~\% + 12~\% + 7.5~\% \pm 7.5~\%)\cdot N_{cut}\quad,
\end{equation}
which finally results in an estimation for the cut efficiency of
channel~3:
\begin{equation}
    \varepsilon_{cut}^{3} = \frac{N_{cut}}{N_{real}} = \frac{1}{(1.195 \pm
    0.075)} = 0.837 \pm 0.053\quad.
\end{equation}

The PMT quantum efficiency function
$\varepsilon_{QE}^{3}(\lambda)$ of channel~3, with a maximum value
of 26.6~\% at 420 nm, is assumed to be known with an uncertainty
of 5~\%, which implies a normalization factor $f_{3} = 1.00 \pm
0.05$. For the collection efficiency $\varepsilon_{col}^{3}$ a
value of $0.880 \pm 0.088$ has been assumed. Using these values
the absolute calibration factor $f_{cal}^{3}$ of the reference
channel~3 can be estimated according to
\begin{equation}
    f_{cal}^{3} = \varepsilon_{cut}^{3}\cdot\varepsilon_{col}^{3}\cdot f_{3} = 0.737 \pm
    0.095\quad,
\end{equation}
with a systematic uncertainty of about 13~\%. By means of
Eq.~(\ref{eq:relCalibration}) the calibration constants of the other channels can be derived relative to $f_{cal}^{3}$. Relative calibration
measurements have been performed once a month over the whole
data-taking period and turned out to be extremely stable~\cite{Waldenmaier:2006}.
The resulting detection efficiencies $\varepsilon_{det} =
f_{cal}\cdot\varepsilon_{\Omega}\cdot\varepsilon_{s}$ for each
nitrogen band in the different filter channels, have been calculated assuming a line spectrum and are listed in Table~\ref{tab:channels}. The corresponding systematic errors are the major source of uncertainty for the absolute determination of the fluorescence yield. Further details about the experiment and the calibration procedures can be found in \cite{Waldenmaier:2006}.
\section{Measurements \& Data Analysis}\label{sec:Analysis}
The data presented in this work were taken from August to November
2005 consisting of about 50 runs in pure nitrogen
($f_{\mathrm{N_{2}}} = 1.00$), dry artificial air
($f_{\mathrm{N_{2}}} = 0.78$, $f_{\mathrm{O_{2}}} = 0.21$,
$f_{\mathrm{Ar}} = 0.01$) and a nitrogen-oxygen mixture
($f_{\mathrm{N_{2}}} = 0.90$, $f_{\mathrm{O_{2}}} = 0.10$) to
reveal the quenching effects of the different air constituents. In
addition several runs with pure nitrogen plus a variable amount of
water vapor have been performed. The measurements were done at
pressures ranging from 2~hPa to 990~hPa at room temperatures
between $\mathrm{15~^{\circ}C}$ and $\mathrm{23~^{\circ}C}$,
varying with the outside weather conditions. In order to
accumulate sufficient statistics, a single run lasted between 12
and 30 hours, depending on the gas mixture and the pressure.
Before each run the chamber was flushed and filled with fresh gas
to avoid aging and contamination effects. Extensive calibration
measurements~\cite{Waldenmaier:2006} were carried out before each
series of runs with a certain gas.
\subsection{Global Minimization Function}\label{sec:minFunction}
Challenging for the data analysis are filter channels containing
more than one nitrogen band, as is illustrated in
Fig.~\ref{fig:filters}. The only channel with a nearly pure
contribution of a single nitrogen transition is channel~4 which is
measuring the prominent 2P(0,0) band at 337.1~nm. Therefore, the
main task for the data analysis is the separation of the different
contributions in the individual filter channels. This can be
achieved to some extend by a global analysis of the complete set
of measurements, making use of the physical relations between the
different nitrogen bands, as discussed in
Section~\ref{sec:NitrogenFluorescence}. Technically this was done
by a constrained $\chi^{2}$-minimization of all datasets, where a
dataset consists of the synchronous data of all filter channels of
a run with a certain gas at a certain pressure. The most universal
constraints directly follow from the general decay
law~(\ref{eq:decaylaw}) and require identical lifetimes
$\tau_{v'}(p,T)$ and intrinsic yields $Y_{v'}^{0}$ for nitrogen
bands emerging from the same initial state $v'$ in all filter
channels of a dataset. Furthermore, the intensity ratios
$R_{v',v''}$ of transitions from the same initial state $v'$ must
have the same values in all datasets and therefore link different
runs with each other. These constraints are always valid since
they do not depend on any assumption about the excitation or
de-excitation mechanism. To further improve the results
additional, model dependent, constraints may be applied such as
the linear pressure dependence of the reciprocal lifetimes as
follows from Eq.~(\ref{eq:recLifetime}). This global approach
directly results in a consistent description of the fluorescence
yield with a minimal set of parameters. Accordingly, the global
minimization function was constructed as a sum over all datasets
$d$, filter channels $c$ and time bins $b$ of the corresponding
time distributions:
\begin{equation}\label{eq:minFunction}
     \chi^{2} = \sum_{d}\sum_{c=1}^{5}\sum_{b}\left[\frac{y_{dcb} -
    \left(O_{dc} + B_{dcb} +
    S_{dcb}\right)}{{\sigma_{y}}_{dcb}}\right]^{2}~~.
\end{equation}
In this expression, $y_{dcb}$ denotes the bin content with
statistical error ${\sigma_{y}}_{dcb} = \sqrt{y_{dcb}}$ and
$O_{dc}$ corresponds to the constant noise offset per bin of
filter channel $c$ in the dataset $d$. The noise offset $O_{dc}$
was individually determined for each time distribution~(see
Fig.~\ref{fig:timeSpec}) by averaging over the bins between 2~ns
and 12~ns in front of the rising edge of the fluorescence signal.
The electron correlated background $B_{dcb}$ was determined and
parametrized by means of background measurements with a
completely evacuated chamber. Its origin was not fully understood, however it is only relevant for low pressure measurements~\cite{Waldenmaier:2006}. Finally the
total fluorescence signal $S_{dcb}$ in each bin $b$ is a sum over
all nitrogen transitions contributing to this channel
\begin{equation}\label{eq:totSignal}
    S_{dcb} = \varepsilon_{DAQ}^{dc}\cdot\sum_{v', v''} \varepsilon_{det}^{cv'v''}\cdot F_{b}(R_{v',v''}\cdot N_{dv'},~ \tau_{dv'})~~,
\end{equation}
where the summation is done over all initial
electronic-vibrational states $v'$ as well as over the final
states $v''$. Note that in the case of contributions from
different electronic band systems the summation over $v'$ includes
the summation over the electronic band systems as well! The
function $F_{b}(R_{v',v''}\cdot N_{dv'},~ \tau_{dv'})$ is the
integral of Eq.~(\ref{eq:TimeDistribution}) over the width of bin
$b$ and depends on the lifetime $\tau_{dv'}$ of the initial state
$v'$ and on the total number of emitted photons by the transition
$v' \rightarrow v''$ which are related to the total amount of
photons $N_{dv'}$ of the main-transition by $R_{v',v''}\cdot
N_{dv'}$. In addition, even if not explicitly quoted in
Eq.~(\ref{eq:totSignal}), $F_{b}$ depends on the time origins
$t_{0}$ and time resolutions $\sigma_{t}$ of each channel which
are kept at the same values in all datasets during the
minimization procedure. In order to retrieve the number of sampled
signals $S_{dcb}$, each integral function $F_{b}$ must be
multiplied by the detection efficiency
$\varepsilon_{det}^{cv'v''}$ of the corresponding transition and
the data acquisition efficiency $\varepsilon_{DAQ}^{dc}$ (duty
cycle) of the filter channel. The detection efficiencies have been
determined by calibration measurements, as explained in the
previous section, and are listed in Table~\ref{tab:channels},
whereas the data acquisition efficiencies $\varepsilon_{DAQ}^{dc}$
are simultaneously measured during a run using the scaler values
and were on the order of 91~\% for pure nitrogen and 96~\% for dry
air~\cite{Waldenmaier:2006}.

The only parameters which are finally varied during the global
minimization procedure thus are the total number of photons of the
main-transitions $N_{dv'}$, the intensity ratios $R_{v',v''}$ and
the lifetimes $\tau_{dv'}$ of the initial states $v'$. If the
linear pressure dependence of the reciprocal lifetimes is taken as
an additional constraint the lifetime $\tau_{dv'}$ in
Eq.~(\ref{eq:totSignal}) is substituted by
relation~(\ref{eq:totDecayConstant}) and the intrinsic lifetime
$\tau_{0}^{v'}$ as well as the various quenching rate constants
$Q_{x}^{v'}(T)$ are varied instead. In this way the consistency of
the parameters is always ensured.

According to the separation of the excitation and de-excitation
processes in Eq.~(\ref{eq:flYieldFinal}) the following data
analysis was subdivided into two steps. First, the de-excitation
parameters such as intrinsic lifetimes $\tau_{0}^{v'}$, intensity
ratios $R_{v',v''}$ as well as the various quenching rate
constants $Q_{x}^{v'}(T)$ for the different air constituents, were
determined using the full usable energy range between 250~keV and
2000~keV to maximize statistics. Afterwards, the obtained values
for the de-excitation parameters were fixed and the intrinsic
fluorescence yields $Y_{v'}^{0}(E)$ of the corresponding
main-transitions were determined in different energy intervals, to
study their energy dependence.
\subsection{Study of De-Excitation Parameters}
The present study concentrates on the eight strongest nitrogen
transitions listed in Table~\ref{tab:channels} which belong to the
three independent vibrational sub-systems 2P(0,$v''$), 2P(1,$v''$)
and 1N(0,$v''$). To obtain stable and reliable results by the
global minimization procedure explained above, at least two
nitrogen transitions from different vibrational sub-systems need
to be measured separately. With the current filter set this was
only possible for the 2P(0,0) transition in channel~4.
Nevertheless, the 2P(1,0) transition in channel~1 can also be
approximately considered as a pure transition since it contains
only small contributions of the 2P(2,1) and 2P(3,2) bands, as can
be seen in Fig.~\ref{fig:filters}.

\begin{figure*}
\centering %
\subfigure[Nitrogen at 800~hPa (smaller time scale!)]{ \centering
\includegraphics[width=0.85\textwidth]{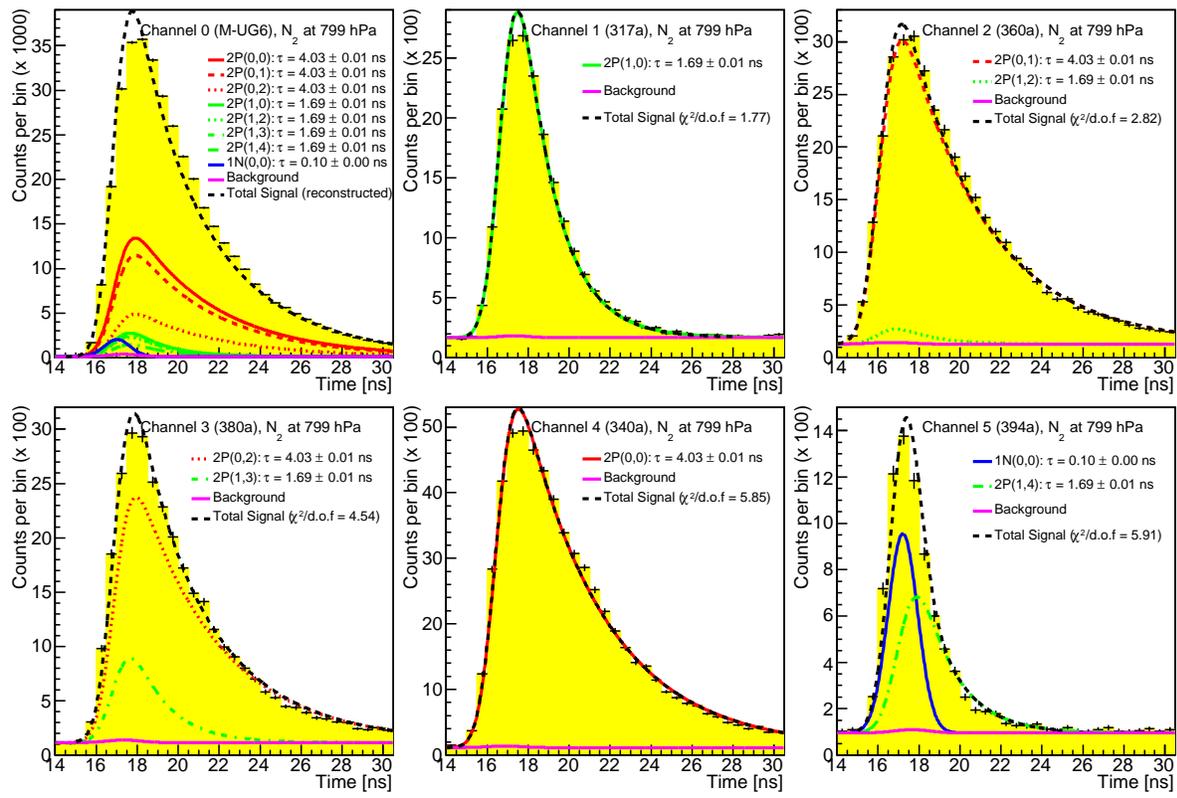}}
\subfigure[Nitrogen at 20~hPa]{ \centering
\includegraphics[width=0.85\textwidth]{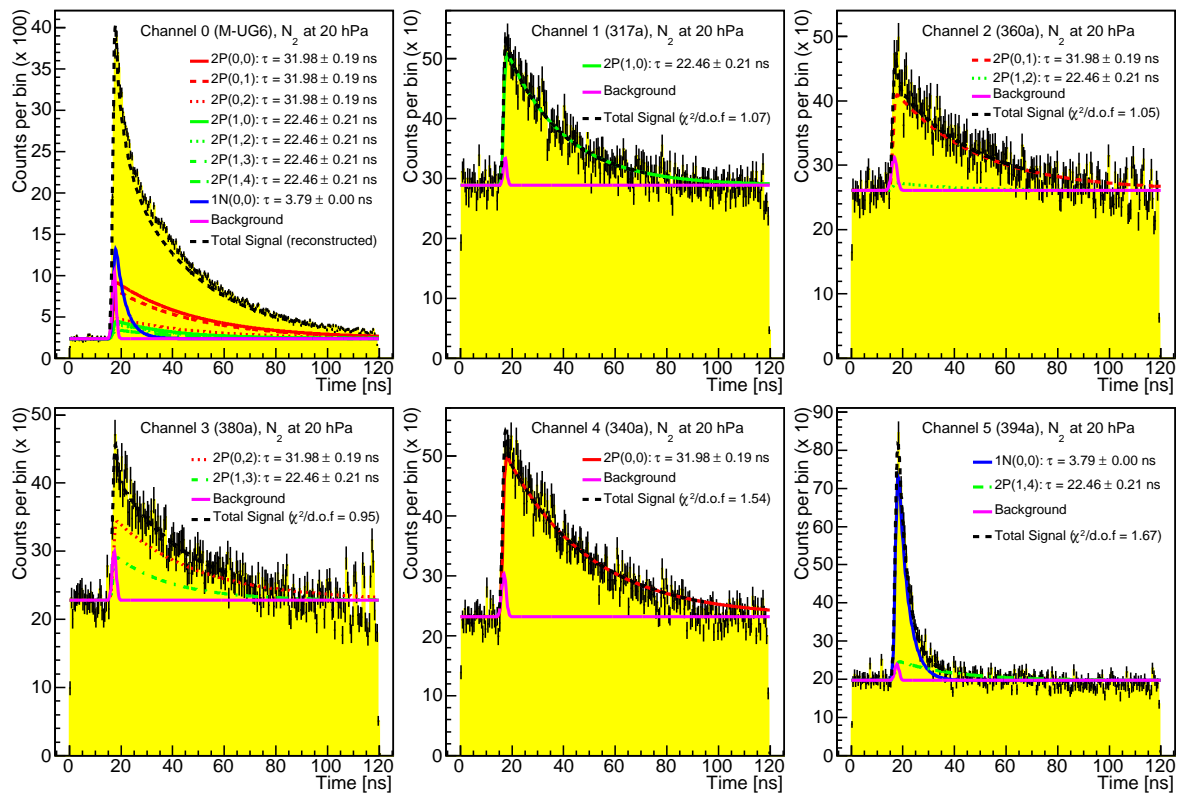}}
\caption{Time spectra in pure nitrogen at 20~hPa and 800~hPa
(different time ranges are displayed in both cases). The signal in
channel~0 is not fitted but has been reconstructed from the signals in
the other channels (see also Fig.~\ref{fig:recSignalMUG6}). For
low pressures, the electron correlated background contributions (pink line) are not
negligible and have been parametrized by an empirical formula~\cite{Waldenmaier:2006}.}\label{fig:timeSpecsNitrogen}
\end{figure*}

\begin{figure*}
\subfigure[Vibrational level $v' = 0$]{ \centering
\includegraphics[width=0.49\textwidth]{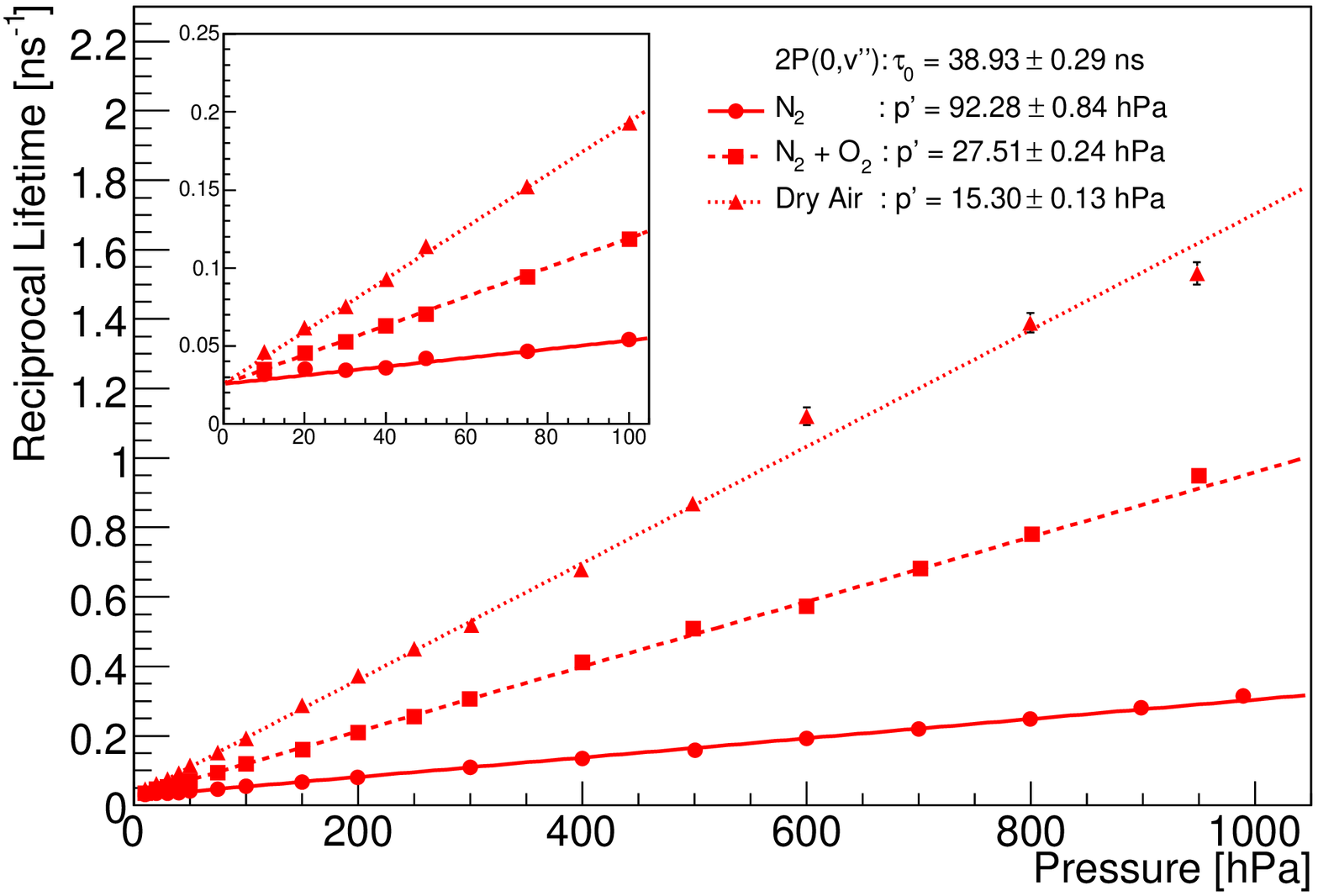}}
\hfill%
\subfigure[Vibrational level $v' = 1$]{ \centering
\includegraphics[width=0.49\textwidth]{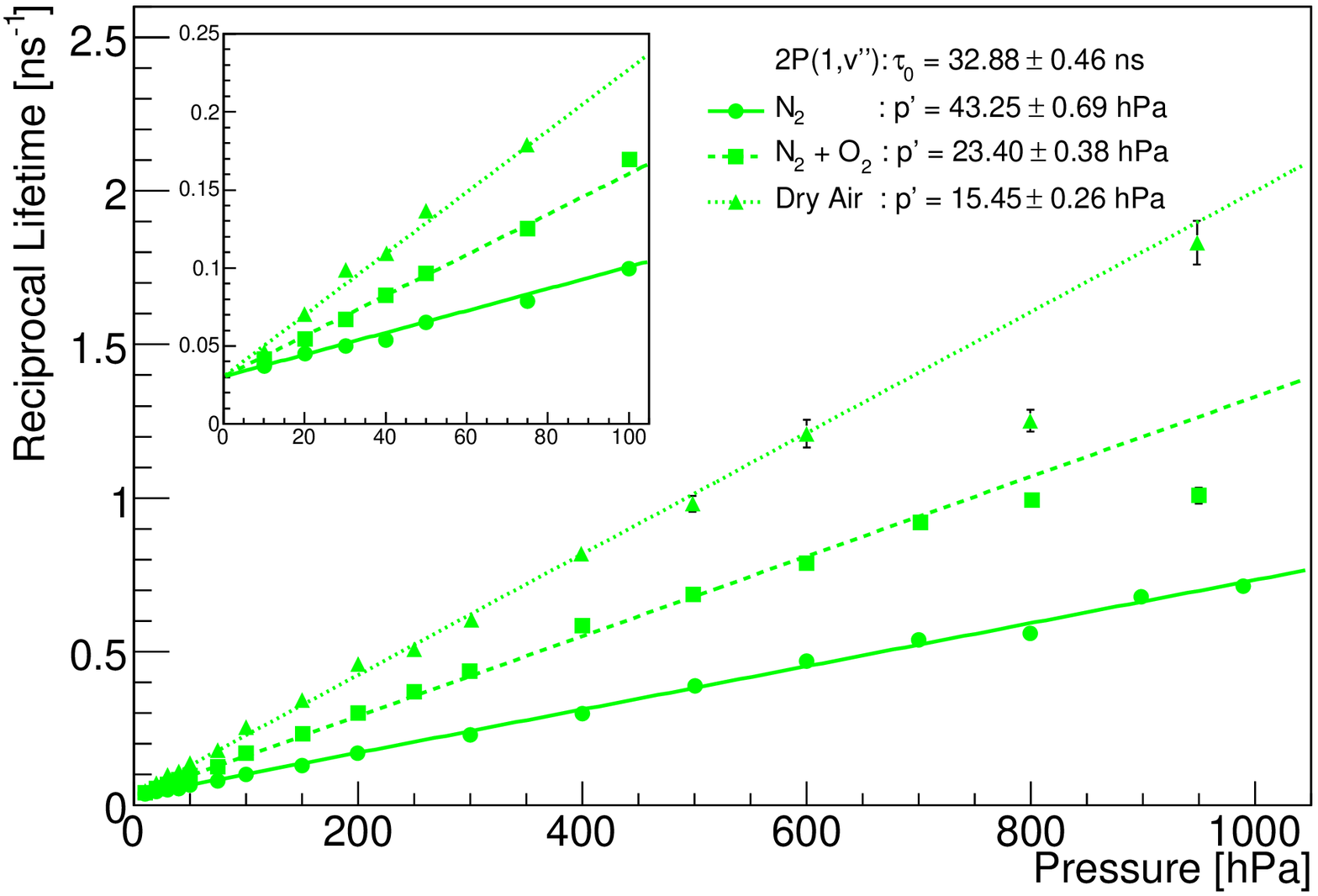}}
\caption{Stern-Vollmer plots of the reciprocal lifetimes for the
vibrational states $v'=0,1$ of the 2P system. The lines are not a
direct fit to the data points but illustrate the pressure
dependence derived from the lifetimes and the quenching rate
constants, which were obtained by a global minimization with a
linear constraint on the reciprocal lifetimes.}
\label{fig:sternVollmer2P}
\end{figure*}

An additional challenge was the measurement of the 1N(0,0)
transition in channel~5, since it is contaminated by several weak
bands of the higher vibrational transitions 2P(3,6), 2P(2,5) and
2P(1,4). Fortunately, transitions of the 1N system have much
longer intrinsic lifetimes and are also more affected by the
collisional quenching than the 2P transitions. Therefore the
evolution of their time spectra with pressure differs much from
the 2P transitions, what helps to separate both band systems from
each other. Since only 2P transitions with $v' = 0,1$ are
considered in this analysis, all the other 2P contaminations in
channel~5 with $v' > 1$ are merged into the 2P(1,4) transition
assuming their pressure dependence to be similar. In this way, the
2P(1,4) transition becomes an effective measure for all 2P
transitions in channel~5 and is expected to be more intense than
the pure 2P(1,4) transition. In order to avoid negative effects on
the results for the 2P transitions in the other filter channels,
the analysis was done in two steps. At first, only the
2P-transitions in channels~1~to~4 were simultaneously studied by
excluding channel~5 from the analysis. Afterwards, just channel~5
was analyzed, keeping the lifetime of the 2P(1,4) transition in
each dataset fixed to the values obtained by the previous analysis
of the 2P system.

An example for the resulting contributions in each filter channel
is given in Fig.~\ref{fig:timeSpecsNitrogen} where the time
spectra of two datasets in pure nitrogen at pressures of 20~hPa
and 800~hPa are shown. Note that for high pressures the time scale
is much smaller than for the low pressure measurements. At 20~hPa
the 1N(0,0) band in channel~5 is clearly dominating whereas at
high pressure the 2P contaminations are of the same order than the
1N(0,0) signal. Channel~0 with the M-UG6 broad band filter is
always excluded from the minimization procedure since it may
contain other contributions from unknown or disregarded
fluorescence transitions. Nevertheless, its signal can be
reconstructed with the information from the other channels
providing a useful crosscheck of the minimization results as is
shown in Section~\ref{sec:compareMUG6}.
\subsubsection{Analysis of the 2P system}
As explained above, only the filter channels~1~to~4 were used to
study the quenching parameters of the 2P system. Furthermore, the
analysis was restricted to measurements at pressures higher or
equal than 10~hPa to avoid systematic effects due to slight gas
impurities. The time spectra of 46 runs in nitrogen, artificial
air, and the nitrogen-oxygen mixture have been simultaneously
fitted by the general minimization
function~(\ref{eq:minFunction}). In the first instance the
minimization function was only constrained by the universal
relations following from the decay law~(\ref{eq:decaylaw}). The
resulting lifetimes for the vibrational levels $v'=0,1$ are
reciprocally plotted versus the pressure in the Stern-Vollmer
plots shown in Fig.~\ref{fig:sternVollmer2P}. For all gases the
data points show a clear linear dependence on the pressure as
expected from Eq.~(\ref{eq:recLifetime}). Some outliers may be
explained by an insufficient separation of different contributions
by the minimization procedure. Such effects can be reduced by
further constraining the lifetime in Eq.~(\ref{eq:totSignal})
through the linear expression~(\ref{eq:totDecayConstant}), which
additionally allows to account for slight temperature variations
between different runs. The results of a second minimization,
using these constraints, are represented by the lines drawn in
Fig.~\ref{fig:sternVollmer2P}. The y-axis intercept is in common
for all lines within a graph, indicating the reciprocal intrinsic
lifetime of the corresponding initial state $v'$. The lines are in
good agreement with the single data points obtained by the
previous minimization without the linear constraint.

This final minimization, with a global reduced $\chi^{2}_{r}$ of
1.77, leads to a consistent set of intrinsic lifetimes
$\tau_{0}^{v'}$, quenching rate constants $Q_{x}^{v'}$ and
intensity ratios $R_{v',v''}$ which are listed in
Tables~\ref{tab:compLifetimes}~-~\ref{tab:compIntRatios} together
with the results of other authors. The value of 32.9~ns for the
intrinsic lifetime of the 2P(1,$v''$) sub-system seems to be too
low compared to the other authors. The reason might be the
contamination of the 2P(1,0) main-transition in channel~1 with
slight contributions of the 2P(2,1) and 2P(3,2) transitions which
cannot be separated by this analysis. This contamination is also
supposed to be the main reason for the deviation of the intensity
ratios of the 2P(1,$v''$) sub-system from the theoretical
expectations of Gilmore et al. in Table~\ref{tab:compIntRatios}.

\subsubsection{Analysis of the 1N system}
\begin{figure}[t]
\includegraphics[width=\linewidth]{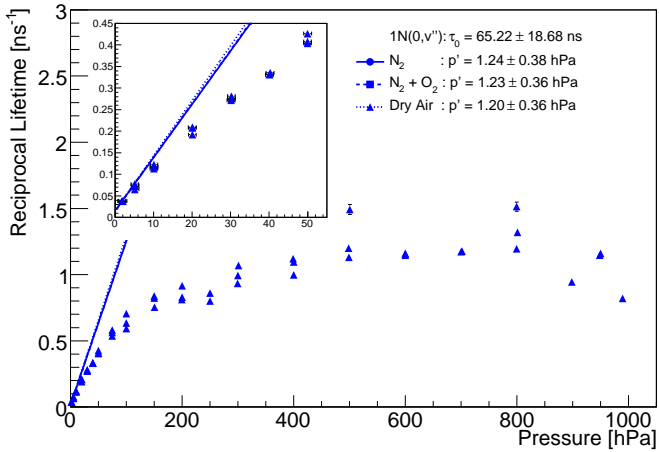}
\caption{Stern-Vollmer plot of the reciprocal lifetimes for the
1N(0,v'') sub-system. As in Fig.~\ref{fig:sternVollmer2P} the
lines correspond to the pressure dependence obtained by a global
minimization with a linear constraint on the reciprocal lifetimes.
The deviation between the single data points and the minimization
result is due to unseparated 2P-contaminations.}
\label{fig:sternVollmer1N}
\end{figure}

The 1N(0,0) transition is the only strong transition of the 1N
system between 300~nm and 400~nm which was studied in this work.
Due to the reasons explained before, the quenching parameters of
this transition were determined by a separate analysis of
channel~5 only. To separate the 1N(0,0) transition from the 2P
contaminations, the lifetime of the 2P(1,4) transition was fixed
in each dataset to the corresponding value taken from the previous
analysis of the 2P system, and only the intensity ratio $R_{1,4}$
was allowed to vary during the global minimization procedure. In
contrast to the 2P system, the pressure range could be extended
down to 2~hPa since the 1N transitions become much stronger at
lower pressures and are also less affected by gas impurities due
to the large self-quenching rate of nitrogen. As for the 2P system
the minimization was done in two steps with increasing levels of
constriction. The first minimization, using only the most general
constraints, results in the individual data points for the
intrinsic lifetimes shown in the Stern-Vollmer plot in
Fig.~\ref{fig:sternVollmer1N}. Compared to the corresponding plots
of the 2P system in Fig.~\ref{fig:sternVollmer2P}, a linear
behavior is hardly visible. This is caused by an insufficient
separation of the 2P contaminations, for pressures larger than
50~hPa. Below this point the data seem to adopt a linear behavior
since the intensity of the 1N(0,0) transition becomes strong
enough to be silhouetted against the 2P contaminations, as can
also be seen in the time distributions of
Fig.~\ref{fig:timeSpecsNitrogen}. Therefore only runs between
2~hPa and 50~hPa were considered in the second minimization step
in which the lifetimes were additionally constrained by the linear
expression~(\ref{eq:totDecayConstant}), as was already done for
the 2P system. This minimization resulted in a global reduced
$\chi^2_{r}$ of 1.26 and a set of de-excitation parameters which
are listed in
Tables~\ref{tab:compLifetimes}~-~\ref{tab:compIntRatios} together
with the values for the 2P system. It must be stressed again that
the intensity ratio of the 2P(1,4) transition in
Table~\ref{tab:compIntRatios} is just an effective measure for all
2P contaminations in channel~5. Therefore its value is expected to
be larger than for the pure 2P(1,4) transition.

The results of this final minimization are represented by the
lines in Fig.~\ref{fig:sternVollmer1N}, which are clearly
deviating from the single data points obtained by the first
minimization without the linear constraint. Therefore, the
separation of the 2P from the 1N contributions in channel~5 only
seems to be possible by using all the known physical relations to
constrain the minimization function. Furthermore, nearly no
difference between the three gas mixtures can be recognized, since
the slopes of the lines turn out to be roughly the same. This
leads to the conclusion that the quenching of the 1N(0,$v''$)
sub-system is mainly due to nitrogen self-quenching, since the
different contributions of oxygen show hardly any effect on the
lifetimes.

\begin{figure}[t]
\includegraphics[width=\linewidth]{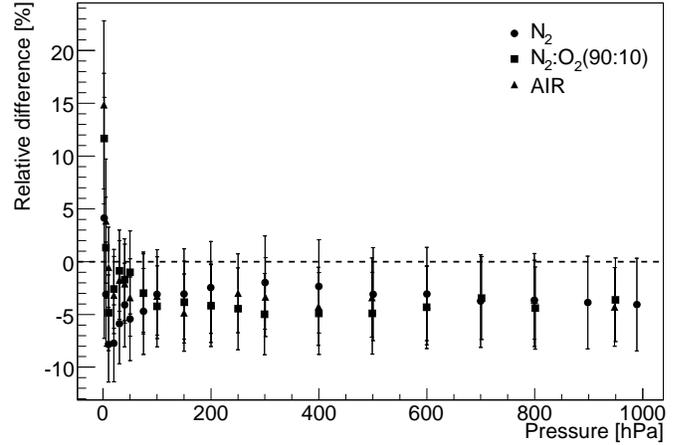}
\caption{Relative differences between the reconstructed and the
measured signal in channel~0 with the M-UG6 filter (systematic
errors are included). From this follows that the contribution of
neglected nitrogen bands is less than 4~\%.}
\label{fig:recSignalMUG6}
\end{figure}

\subsubsection{Comparison to M-UG6 Data}\label{sec:compareMUG6}
The de-excitation parameters determined in the last two sections
completely describe the pressure dependence of the relative
intensities for the investigated nitrogen transitions. To verify
their applicability on the whole atmospheric pressure range, the
time spectra of channels~1 to 5 of the complete set of
measurements were fitted again with the de-excitation parameters
fixed to the values listed in
Tables~\ref{tab:compLifetimes}~-~\ref{tab:compIntRatios}. Only the
intensities of the main-transitions were allowed to vary freely in
each dataset. Afterwards, the fluorescence signal in channel~0
with the M-UG6 broad band filter was calculated using the fitted
intensities and the detection efficiencies (see
Table~\ref{tab:channels}) of each transition. An example for the
reconstructed time distributions in channel~0 is given by the
upper left graphs of Fig.~\ref{fig:timeSpecsNitrogen} for pure
nitrogen at two different pressures. The yellow histograms
represent the measured time distributions and the calculated
contributions of each nitrogen transition, as well as their sum,
are plotted on top of the measured distribution. For both
pressures, the signal in channel~0 could be almost completely
recovered by the information from the other channels. This worked
also for the other gas mixtures and pressures, as is shown in
Fig.~\ref{fig:recSignalMUG6}, where the relative differences
between the reconstructed and measured integrated fluorescence
signals in channel~0 are plotted versus the pressure. The
systematic uncertainties are already included in the error bars
and the large fluctuations of the data points below 10~hPa can be
explained by the limited statistics for low pressure runs. For
pressures larger than 10~hPa the data points stabilize and the
reconstructed signals are about 4~\% smaller, but still compatible
with the measured signals. However, a small excess of the measured
signal in channel~0 is expected, since this channel is also
measuring transitions which were neglected in the present
analysis. From this follows that the contribution of disregarded
transitions to the total nitrogen fluorescence spectrum in the
M-UG6 filter range is less than 4~\%.
\begin{figure*}[t]
\subfigure[2P system]{ \centering
\includegraphics[width=0.49\textwidth]{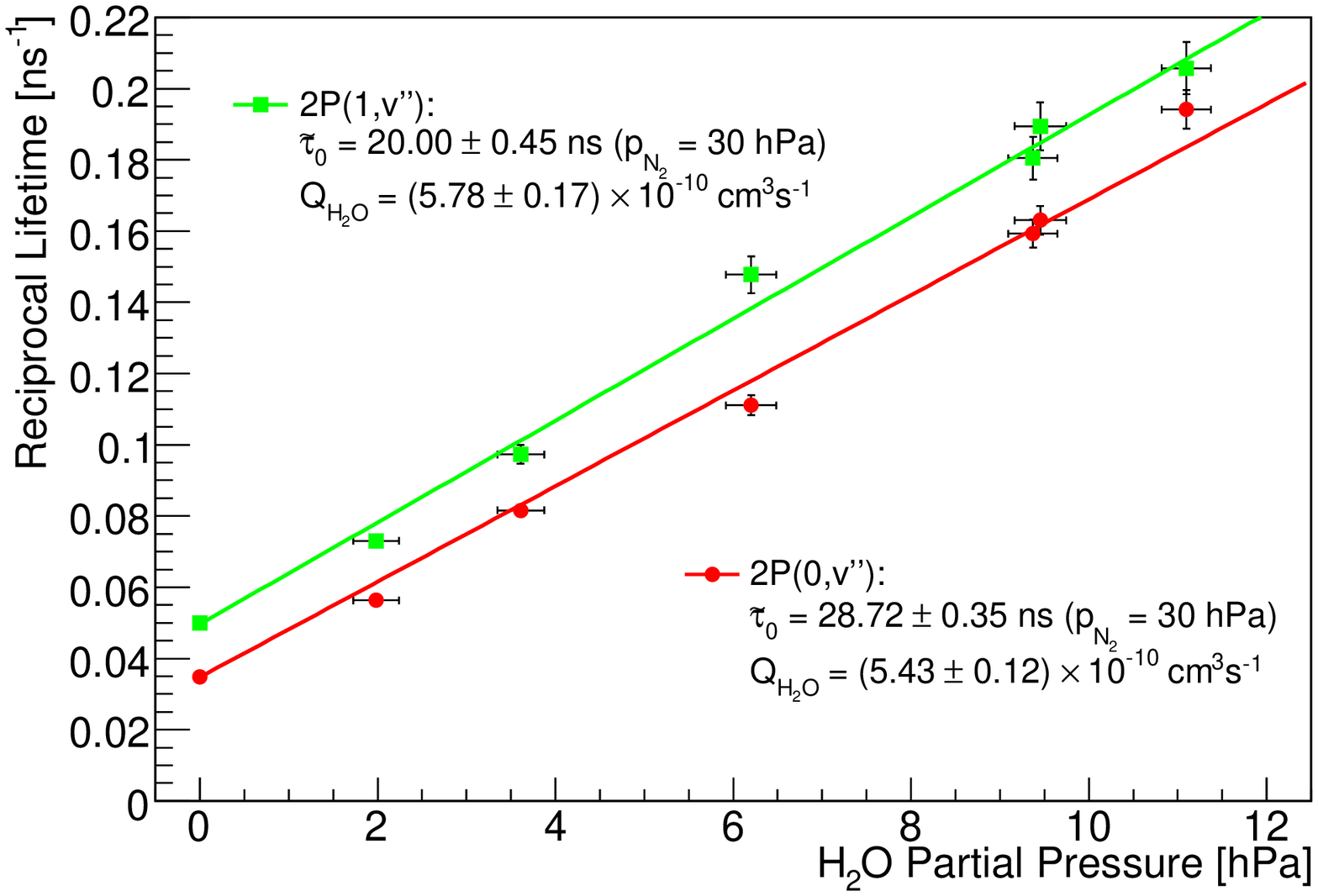}}
\hfill%
\subfigure[1N system]{ \centering
\includegraphics[width=0.49\textwidth]{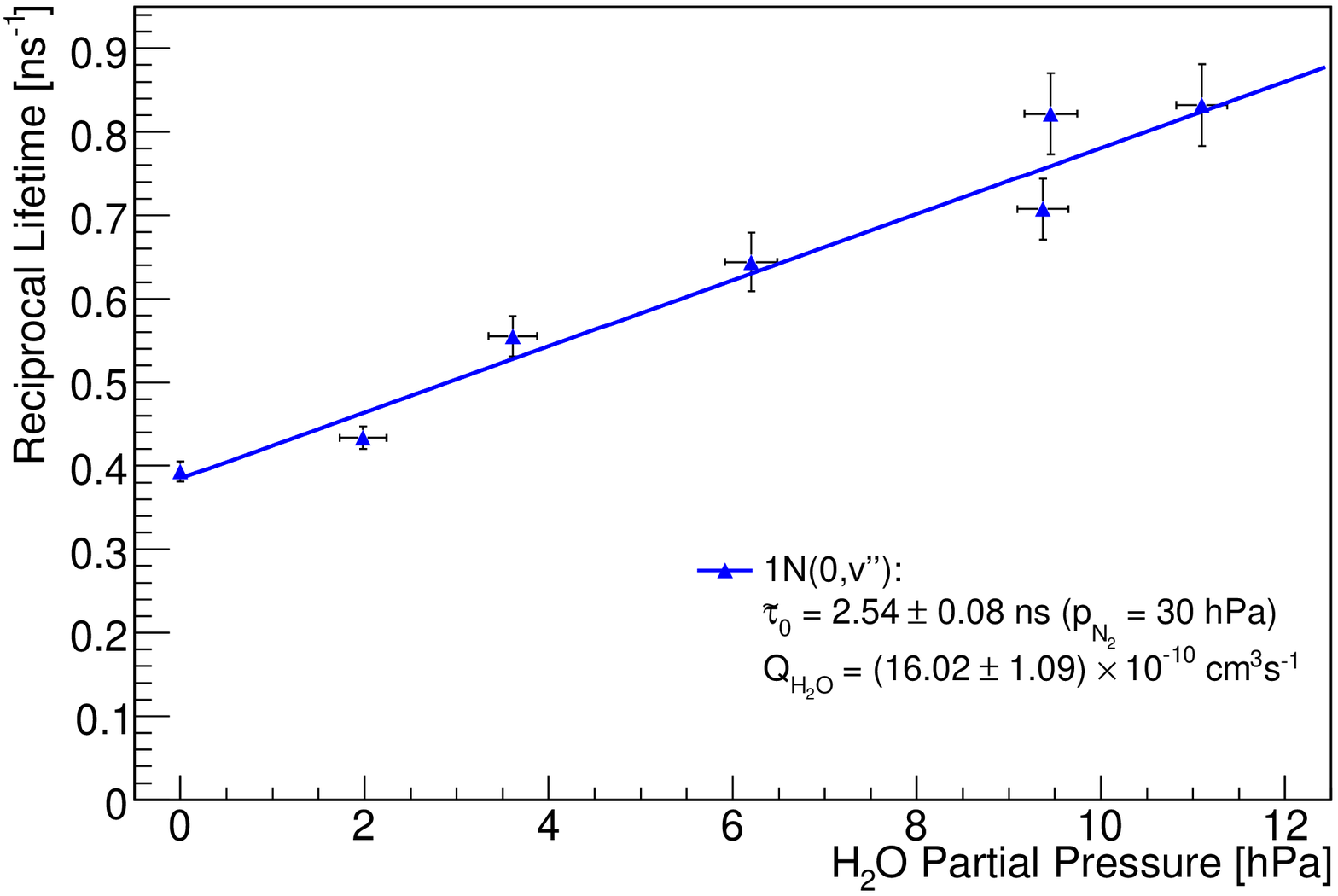}}
\caption{Reciprocal lifetimes in humid nitrogen versus the water
vapor partial pressure. The nitrogen partial pressure was always
kept at 30~hPa. The data points at $p_{\mathrm{H_{2}O}}=0$
correspond to the reciprocal lifetimes $1/\tilde{\tau_{0}}$ in dry
nitrogen at a pressure of 30~hPa.} \label{fig:sternVollmerH2O}
\end{figure*}
\subsubsection{The Effect of Water Vapor}
The influence of water vapor on the nitrogen quenching was studied
by adding different concentrations of water vapor to pure nitrogen
at a (partial-)~pressure of
$p_{\mathrm{N_{2}}}~=~30~\mathrm{hPa}$. The water vapor
concentration was monitored by a calibrated humidity
probe\footnote{AHLBORN FH A646-E7C} which was measuring the
relative humidity with an absolute systematic uncertainty of
1.5~\%~rH. The water vapor partial pressure $p_{\mathrm{H_{2}O}}$
was derived from the relative humidity according to the magnus
formula~\cite{Waldenmaier:2006,Murray:1967}. Six runs were
performed at room temperatures between $\mathrm{15~^{\circ}C}$ and
$\mathrm{17~^{\circ}C}$ and relative humidities ranging from
10~\%~rH to 60~\%~rH. Each run lasted for roughly 30 hours but
only data of the last $\sim$~20 hours was used for the analysis,
in order to ensure the water vapor concentration to be in thermal
equilibrium with the surrounding chamber walls.

As before, the data analysis was done by simultaneously fitting
the time spectra of all datasets with the minimization
function~(\ref{eq:minFunction}), by applying only the most general
constraints and by fixing the intensity ratios to the values
obtained from the previous analysis listed in
Table~\ref{tab:compIntRatios}. The global reduced $\chi^{2}_{r}$
of this minimization was 1.09 and the resulting reciprocal
lifetimes for the three vibrational sub-systems are drawn versus
the water vapor partial pressure $p_{\mathrm{H_{2}O}}$ in
Fig.~\ref{fig:sternVollmerH2O}. The data points at
$p_{\mathrm{H_{2}O}}=0$ correspond to the results of the previous
analysis for dry nitrogen at a pressure of 30~hPa. According to
Eq.~(\ref{eq:totDecayConstant}) the water vapor quenching rate
constants $Q^{v'}_{\mathrm{H_{2}O}}$ were determined through a fit
of the linear expression
\begin{equation}
\frac{1}{\tau_{v'}} = \underbrace{\frac{1}{\tau^{v'}_{0}} +
\frac{p_{\mathrm{N_{2}}}}{kT}\cdot
Q^{v'}_{\mathrm{N_{2}}}}_{=:~1/\tilde{\tau}_{0}} +
\frac{p_{\mathrm{H_{2}O}}}{kT}\cdot Q^{v'}_{\mathrm{H_{2}O}}
\end{equation}
to the data points in Fig.~\ref{fig:sternVollmerH2O}, resulting in
the values listed in Table~\ref{tab:compQuenchRates}. For the 2P
system the water vapor quenching rate constants were roughly two
times larger than the corresponding values for oxygen, whereas the
quenching of the 1N system by water vapor is roughly three times
stronger than by nitrogen or oxygen. This large value may be
explained by the polar nature of the $\mathrm{H_{2}O}$ and the
$\mathrm{N_{2}^{+}}$ molecules. Despite of the large values for
the water vapor quenching rate constants the effect on the
nitrogen quenching in the atmosphere is expected to be rather
small since the water vapor concentration in air usually is much
lower than the nitrogen or oxygen concentrations.

\begin{table*}
\hspace{0.11\textwidth}
\begin{minipage}{0.89\textwidth}
\caption{Intrinsic lifetimes $\tau_{0}^{v'}$ in [ns] for the
initial states of the investigated vibrational sub-systems.
Only\newline statistical errors are quoted.}
\label{tab:compLifetimes}
\begin{tabular}{ccccl}
    \hline
    ~~~~~~~~~2P($v'=0$)~~~~~~~~~ & ~~~~~~~~~2P($v'=1$)~~~~~~~~~ & ~~~~~~~~~1N($v'=0$)~~~~~~~~~& ~~~~~~~~ & Reference\\
    \hline
    $38.9 \pm 0.3 $ & $32.9 \pm 0.5$ & $65.2 \pm 18.7$ &~& this work\\
    $42.0 \pm 2.0$ & $41.0 \pm 3.0$ & - &~& Pancheshnyi et al. \cite{Pancheshnyi:2000}\\
    $41.7 \pm 1.4$ & $41.7 \pm 2.1$ & - &~& Morozov et al. \cite{Morozov:2005}\\
    $44.5 \pm 6.0$ & $49.3$ & $65.8 \pm 3.5$ &~& Bunner \cite{Bunner:1967}\\
    $37.1^{*}$  & $37.5^{*}$ & $62.3^{*}$ &~& Gilmore et al. \cite{Gilmore:1992}\\
    \hline
    {\small$^{*}$~Theoretical values}\\
\end{tabular}
\vspace{10mm}
\caption{Quenching rate constants $Q_{x}^{v'}$ in
[$\mathrm{10^{-10} cm^{3}s^{-1}}$] for the initial states of the
investigated vibrational\newline sub-systems, given for $T=293
K~(\sim\mathrm{20^{\circ}C}$). Only statistical errors are
quoted.}\label{tab:compQuenchRates}
\begin{tabular}{cccccl}
    \hline
    Molecule & ~~~~~~~2P($v'=0$)~~~~~~~ & ~~~~~~~2P($v'=1$)~~~~~~~ & ~~~~~~~1N($v'=0$)~~~~~~~ & ~~~~~~~ & Reference\\
    \hline
    $\mathrm{N_{2}}$ & $0.11 \pm 0.00$ & $0.29 \pm 0.00$ & $5.00 \pm 0.17$ &~& this work\\
     ~ & $0.13 \pm 0.02$ & $0.29 \pm 0.03$ & - &~& Pancheshnyi et al.~\cite{Pancheshnyi:2000}\\
     ~ & $0.12 \pm 0.01$ & $0.25 \pm 0.01$ & - &~& Morozov et al.~\cite{Morozov:2005}\\
     ~ & $0.10 \pm 0.01$ & $0.22 \pm 0.03$ & $3.80 \pm 0.26$ &~& Nagano et al.~\cite{Nagano:2003}\\
     ~ & $0.07$  & $0.23$ & $2.90$ &~& Bunner \cite{Bunner:1967}\\
    \hline
    $\mathrm{O_{2}}$ & $2.76 \pm 0.01$ & $2.70 \pm 0.03$ & $5.24 \pm 0.79$ &~& this work\\
     ~ & $3.00 \pm 0.30$ & $3.10 \pm 0.30$ & - &~& Pancheshnyi et al.~\cite{Pancheshnyi:2000}\\
     ~ & $2.62 \pm 0.19$ & $2.77 \pm 0.45$ & $2.39 \pm 0.40$ &~& Nagano et al.~\cite{Nagano:2003}\\
     ~ & $1.35$ & - & $8.38$ &~& Bunner \cite{Bunner:1967}\\
    \hline
    $\mathrm{H_{2}O}$ & $5.43 \pm 0.12$ & $5.78 \pm 0.17$ & $16.02 \pm 1.09$ &~& this work\\
     ~ & $3.90 \pm 0.40$ & $3.70 \pm 0.40$ & - &~& Pancheshnyi et al.~\cite{Pancheshnyi:2000}\\
     ~ & $7.10 \pm 0.70$ & $6.70 \pm 0.70$ & - &~& Morozov et al.~\cite{Morozov:2005}\\
    \hline
\end{tabular}
\vspace{10mm}
\caption{Intensity ratios $R_{v',v''}$ of the investigated
transitions. The values of Bunner have been derived from
the\newline fluorescence efficiencies quoted in his work. The
second error corresponds to systematic uncertainties\newline of
the detection efficiencies.}\label{tab:compIntRatios}
\begin{tabular}{cccccc}
    \hline
    Transition & ~~~~~~~~~~~~~~~this work~~~~~~~~~~~~~~~ & ~~~~Bunner~\cite{Bunner:1967}~~~ & ~~~~Fons et al.~\cite{Fons:1996}~~~ & ~~~~Gilmore et al.~\cite{Gilmore:1992}~~~\\
    \hline
    2P(0,0)$^{a}$ & $1.00 \pm 0.00 \pm 0.00$ & $1.00$ & $1.00$ & $1.00^{*}$\\
    2P(0,1)     & $0.69 \pm 0.00 \pm 0.06$ & $0.75$ & $0.63$ & $0.67^{*}$\\
    2P(0,2)     & $0.29 \pm 0.00 \pm 0.02$ & $0.26$ & $0.25$ & $0.27^{*}$\\
    \hline
    2P(1,0)$^{a}$ & $1.00 \pm 0.00 \pm 0.00$ & $1.00$ & $1.00$ & $1.00^{*}$\\
    2P(1,2)     & $(0.33 \pm 0.03 \pm 0.04)^{b}$ & $0.58$ & $0.45$ & $0.47^{*}$\\
    2P(1,3)     & $(0.34 \pm 0.01 \pm 0.03)^{b}$ & $0.54$ & $0.43$ & $0.41^{*}$\\
    2P(1,4)     & $(0.46 \pm 0.02 \pm 0.04)^{c}$ & $0.32$ & $0.16$ & $0.20^{*}$\\
    \hline
\end{tabular}\\
\begin{tabular}{l}
$^{*}$~Theoretical values.\\
$^{a}$~Main-transitions (Ratio equals to one per definition).\\
$^{b}$~Biased value due to contamination of 2P(1,0) band.\\
$^{c}$~Biased value due to contamination of 2P(1,0) and 2P(1,4) bands.\\
\end{tabular}
\end{minipage}
\end{table*}
\subsection{Study of Energy Dependence}
According to Eq.~(\ref{eq:flYieldFinal}) the total number of
photons $N_{v',k}$, emitted by a main-transitions $v' \rightarrow
k$, is related to the intrinsic fluorescence yield
$Y^{0}_{v',k}(E)$ through the expression
\begin{equation}\label{eq:intensity}
    N_{v',k}(p,T,E) = \frac{\tau_{v'}(p,T)}{\tau_{0}^{v'}}\cdot
    Y^{0}_{v',k}(E)\cdot E_{dep}(E)~.
\end{equation}
The lifetime $\tau_{v'}(p,T)$ is only affected by the
de-excitation parameters, which are known from the analysis
described before. Therefore the last missing quantity for the
determination of the intrinsic fluorescence yield is the deposited
energy $E_{dep}(E)$ in the chamber volume which was obtained from
Monte-Carlo studies described below.
\subsubsection{Energy deposit in Chamber}\label{sec:EnergyDeposit}
A precise knowledge of the energy deposit in the chamber is
crucial for the correct determination of the fluorescence yield.
In confined volumes, the energy deposit is not necessarily equal
to the ionization energy loss, as it follows from the Bethe-Bloch
formula, since secondary delta electrons may carry away part of
the energy. Furthermore, additional secondary electrons emerging
from collisions of the primary electrons with the collimator
walls, or electrons repulsing out of the scintillator, may even
increase the effective energy deposit. Therefore the energy
deposit in the chamber is expected to deviate from the ionization
energy loss. Energy losses due to Bremsstrahlung in the gas are
negligible at these energies.

\begin{figure}[t]
\includegraphics[width=\linewidth]{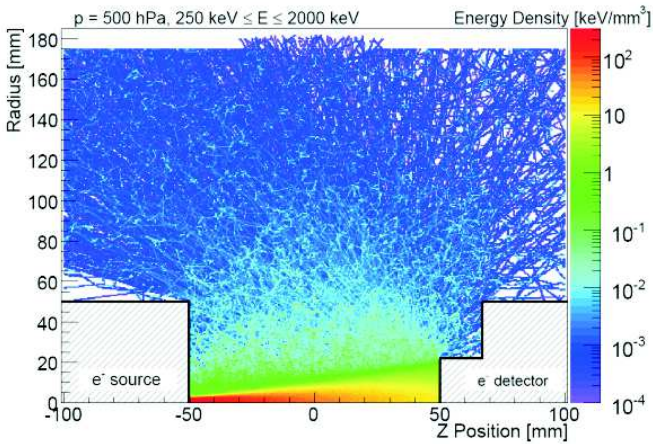}
\caption{GEANT4 simulation of the spatial distribution of the
deposited energy density within the AirLight chamber at a pressure
of 500 hPa, for detected electron energies between 250~keV and
2000~keV.} \label{fig:simEDepMap}
\end{figure}

To determine the mean energy deposit in the chamber of an electron
which was detected with a certain energy in the scintillator,
detailed GEANT4 simulations have been
performed~\cite{Waldenmaier:2006}. The simulations were done using
GEANT4~7.1~\cite{Agostinelli:2002} and the low energy extension
for the electromagnetic processes. The simulated $^{90}$Sr
electron energy spectra were convoluted with the energy resolution
function of the scintillation detector and compared to the
measured energy spectra. The "facrange" parameter of the multiple
scattering model, which affects the penetration depth of the
electrons into a new material, had to be tuned to a value of 0.01.
This significantly improved the treatment of the electron
backscattering within the lead collimator which has to be taken
into account in order to produce realistic energy
spectra~\cite{Waldenmaier:2006}. After the simulation was tested
and verified with real electron spectra, several simulations for
dry air, at all pressures applied to the real measurements, were
performed.

The simulation directly generated two-dimensional histograms of
the deposited energy in each volume element of the chamber, with a
spatial resolution of $\mathrm{1~mm^{3}}$. Each electron, no
matter if it is the primary or a secondary electron, contributes
to the energy deposition. An example for such an energy deposit
map is given in Fig.~\ref{fig:simEDepMap} for detected electron
energies ranging from 250~keV to 2000~keV. If the number of
emitted fluorescence photons is proportional to the deposited
energy, Fig.~\ref{fig:simEDepMap} can be reinterpreted as the
spatial distribution of the fluorescence emissions in the chamber.
This was done to determine the angular acceptance
$\varepsilon_{\Omega}$ of the PMTs~\cite{Waldenmaier:2006} which
was used to calculate the detection efficiencies in
Table~\ref{tab:channels}.

\begin{figure}[t]
\includegraphics[width=\linewidth]{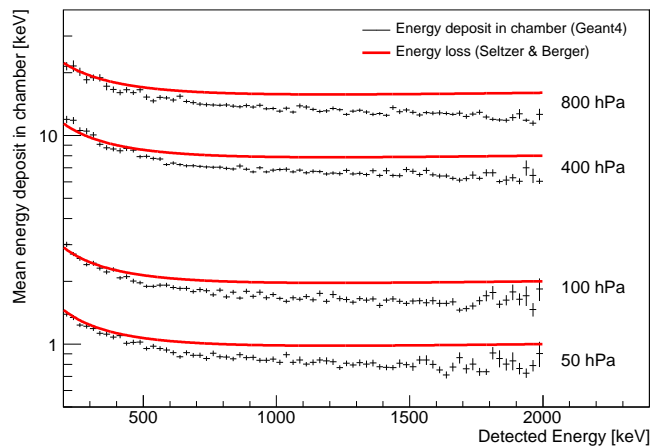}
\caption{Mean deposited energy per electron in the chamber versus
the detected energy in the scintillator for various pressures. The
red lines correspond to the ionization energy loss in dry air
given by the Seltzer \& Berger formula~\cite{Seltzer:1982}.}
\label{fig:eDepEsEdet}
\end{figure}

In order to study the energy dependence of the energy deposition,
several energy deposit maps for successive, 25~keV wide energy
intervals were generated. Each map was integrated over the volume
of the chamber and divided by the number of detected electrons to
derive the mean energy deposit in the chamber per detected
electron. In Fig.~\ref{fig:eDepEsEdet} the mean energy deposit is
plotted for several pressures versus the detected electron energy
in the scintillator. The comparison to the Seltzer~\&~Berger
description~\cite{Seltzer:1982} for the mean ionization energy
loss reveals a 15~\% to 20~\% decrease for electrons detected with
energies larger than 500~keV. Below this point the deposited
energy seems to agree with the ionization energy loss suggesting
most of the secondary electrons to be stopped within the gas
volume of the chamber.
\subsubsection{Intrinsic Yields}
To determine the intrinsic yields $Y_{v',k}^{0}(E)$ and to study
their energy dependence the measured data were divided into seven
sub-samples of 250~keV energy intervals covering the range between
250~keV and 2000~keV. Each sub-sample was fitted separately
according to the minimization procedure explained in
Section~\ref{sec:minFunction}. During the minimization all the
de-excitation parameters were fixed to the values listed in
Tables~\ref{tab:compLifetimes}~-~\ref{tab:compIntRatios} and only
the total number of photons $N_{v',k}$ emitted by the
main-transitions was allowed to vary freely. Afterwards the
intrinsic fluorescence yields $Y_{v',k}^{0}(E)$ in each energy
interval were evaluated by solving Eq.~(\ref{eq:intensity}) using
the total number of photons $N_{v',k}$ from the fit and the
deposited energy, which was calculated from the total number of
detected electrons multiplied by the mean deposited energy per
electron from the simulation studies explained above.

\begin{figure}[t]
\includegraphics[width=\linewidth]{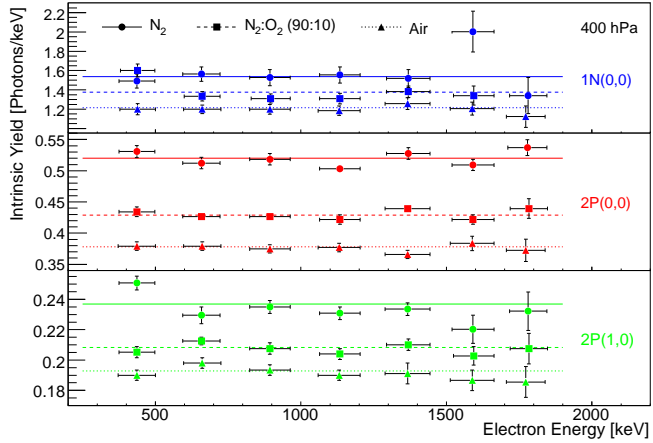}
\caption{Energy dependence of the intrinsic fluorescence yields at
a pressure of 400~hPa. The horizontal lines are not a direct fit
to the data points, but represent the minimization results using
the complete energy range.} \label{fig:intrYield400}
\end{figure}

An example for a medium pressure of 400~hPa is given in
Fig.~\ref{fig:intrYield400} where the intrinsic fluorescence
yields of the three main-transitions are plotted versus the
electron energies. The relative differences between the intrinsic
yields in different nitrogen mixtures seem roughly to scale with
the different nitrogen concentrations. Furthermore the intrinsic
fluorescence yield appears to be constant over the whole energy
range which has been studied. Some fluctuations at high energies
can be explained by the reduced statistics due to the
$\beta$-shape of the electron energy spectrum. In particular
between 250~keV and 1000~keV, where the ionization energy loss
varies about 30~\%, there is no notable change of the intrinsic
fluorescence yields. Therefore, the above analysis was repeated
using the whole usable energy range between 250~keV and 2000~keV.
The results are represented by the horizontal lines in
Fig.~\ref{fig:intrYield400} and agree well with the corresponding
data points of the single energy intervals.

\begin{figure}[t]
\includegraphics[width=\linewidth]{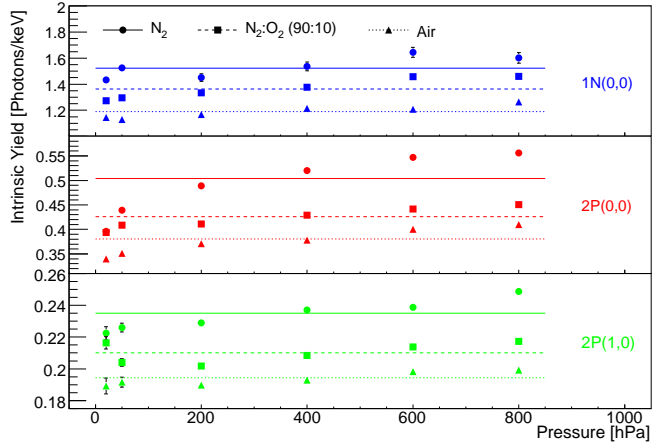}
\caption{Pressure dependence of the intrinsic fluorescence yield
which presumably is due to experimental deficiencies. The
horizontal lines correspond to the values quoted in
Table~\ref{tab:compIntrYields} and denote the weighted average of
the data points.} \label{fig:intrYieldVsPressure}
\end{figure}

\begin{table}[b]
\caption{Systematic uncertainties due to the observed pressure dependence of the intrinsic fluorescence yield.}\label{tab:perrors}
\begin{tabular}{lccc}
\hline
Gas~~~~~~~~~~~~~~~~~ & ~~~~~2P(0,0)~~~~~ & ~~~~~2P(1,0)~~~~~ & ~~~~~1N(0,0)~~~~~~~\\
\hline
$\mathrm{N_{2}}$ & 12~\% & 4~\% & 5~\%\\
$\mathrm{N_{2}:O_{2}}$~(90:10) & 5~\% & 3~\% & 5~\%\\
Dry Air & 7~\% & 2~\% & 4~\%\\
\hline
\end{tabular}
\end{table}
To verify the results, this procedure was repeated for several
pressures even if a pressure dependence was not expected from the
model described in Section~\ref{sec:NitrogenFluorescence}. It
turned out that the intrinsic fluorescence yield apparently
decreases for lower pressures as is shown in
Fig.~\ref{fig:intrYieldVsPressure}. The effect is strongest for
pure nitrogen and seems to decrease with increasing oxygen
concentrations in the other gas mixtures. It could not be
clarified if this behavior is a real physical effect or if it is
due to experimental issues. Therefore, the final intrinsic
fluorescence yield values in Table~\ref{tab:compIntrYields} have
been calculated as the weighted average over the data points at
different pressures, and are represented by the horizontal lines
in Fig.~\ref{fig:intrYieldVsPressure}. The corresponding
systematic uncertainties have been determined as the standard
deviation of the single data points to the averaged value and are
listed in Table~\ref{tab:perrors}. These uncertainties have been
quadratically added to the uncertainties of the detection
efficiencies in Table~\ref{tab:channels} to determine the
systematic errors quoted in Table~\ref{tab:compIntrYields}.
Accordingly the final systematic uncertainties for the individual
nitrogen bands in air settle between 14~\% and 16~\%.

\begin{table*}
\caption{Absolute values of the intrinsic fluorescence yields
$Y^{0}_{v',k}$ in [photons/keV] for the investigated
main-transitions $v' \rightarrow k$. The values of this work are
valid for electron energies ranging from 250~keV to 2000~keV. The
values of Nagano et al. have been converted from his
$\Phi^{0}$-values according to $Y^{0} =
\frac{\lambda}{hc}\Phi^{0}$. The second error corresponds to
systematic uncertainties.} \label{tab:compIntrYields}
\begin{tabular}{lcccl}
    \hline
    Gas~~~~~~~~~~~~~~~~~~ & ~~~~~~~~~~~~~~~2P(0,0)~~~~~~~~~~~~~~~ & ~~~~~~~~~~~~~~~2P(1,0)~~~~~~~~~~~~~~~ & ~~~~~~~~~~~~~~~1N(0,0)~~~~~~~~~~~~~~~ & ~~~Reference~~~~~~~~~~~~~~~ \\
    \hline
    $\mathrm{N_{2}}$ & $0.513 \pm 0.001 \pm 0.096$ & $0.236 \pm 0.001 \pm 0.034$ & $1.513 \pm 0.014 \pm 0.223$ & ~~~this work\\
    ~ & $0.318 \pm 0.005 \pm 0.041$  & $0.129 \pm 0.007 \pm 0.017$ & $0.397 \pm 0.028 \pm 0.052$ & ~~~Nagano et al.~\cite{Nagano:2004}\\
    \hline
    $\mathrm{N_{2}}:\mathrm{O_{2}}~(90:10)$ & $0.429 \pm 0.001 \pm 0.065$ & $0.210 \pm 0.001 \pm 0.030$ & $1.360 \pm 0.009 \pm 0.200$ & ~~~this work\\
    ~ & - & - & - & ~~~Nagano et al.\\
    \hline
    Dry air & $0.384 \pm 0.001 \pm 0.061$ & $0.195 \pm 0.001 \pm 0.028$ & $1.190 \pm 0.008 \pm 0.171$ & ~~~this work\\
    ~       & $0.272 \pm 0.007 \pm 0.035$ & $0.122 \pm 0.007 \pm 0.016$ & $0.303 \pm 0.012 \pm 0.039$ & ~~~Nagano et al.~\cite{Nagano:2004}\\
    \hline
\end{tabular}
\end{table*}
\section{Results and Discussion}\label{sec:results}
At the beginning of this paper an alternative approach to
parametrize the nitrogen fluorescence process in air was
introduced. This approach accounts for all physical relations
between the different nitrogen emission bands and consistently
describes the whole process with a minimal set of parameters. The
excitation and de-excitation processes are completely separated by
this model and the corresponding parameters have a clear physical
meaning. The fluorescence yield in this work is defined in terms
of photons per deposited energy. This definition is much closer
related to the experimental quantities than the traditional
definition in terms of photons per meter track length and
additionally allows to account for effects of escaping delta
electrons in small volumes. The latter issue especially is
important for the absolute determination of the fluorescence yield
in small laboratory experiments.

The AirLight experiment synchronously measured the eight strongest
nitrogen emission bands between 300~nm and 400~nm for pressures
ranging from 2~hPa to 990~hPa with a systematic uncertainty of
about 15~\%. Furthermore the influence of water vapor on the
nitrogen quenching has been studied. The experimental data was
analyzed according to the model mentioned before. In the
investigated energy range between 250~keV to 2000~keV the
fluorescence yield was shown to be independent from the energy of
the ionizing electrons.

\subsection{Usage of the Model}
To calculate the fluorescence yield $Y_{v',v''}(p,T)$ for a
certain nitrogen transition $v' \rightarrow v''$ at a given
temperature~$T$ and pressure~$p$ the formula
\begin{equation}\label{eq:FinalYieldReminder}
Y_{v',v''}(p,T) = Y_{v',k}^{0}\cdot R_{v',v''} \cdot
\frac{\tau_{v'}(p,T)}{{\tau_{0}}_{{v'}}}
\end{equation}
applies, where the value of the associated intrinsic fluorescence
yield $Y_{v',k}^{0}$ is listed in Table~\ref{tab:compIntrYields}
and the values for the intensity ratio $R_{v',v''}$ and the
intrinsic lifetime ${\tau_{0}}_{{v'}}$ are tabulated in
Table~\ref{tab:compIntRatios} and Table~\ref{tab:compLifetimes} compared to the values of other authors. In this context it should also be referred to the results of the AIRFLY experiment~\cite{Ave:2007} which have recently been published. However, it is strongly advised not to mix the intensity ratios and
intrinsic yields from different authors since both quantities are
correlated to each other. Even if some intensity ratios of this
work, especially for the 2P(1,$v''$) sub-system, deviate from the
values of other authors, they still yield reasonable absolute
intensities if they are used together with their associated intrinsic yield values.

The pressure and temperature dependence of the lifetime
$\tau_{v'}(p,T)$ in Eq.~(\ref{eq:FinalYieldReminder}) can be
parametrized according to Eq.~(\ref{eq:totDecayConstant}) and result in the following expression for the reciprocal lifetime
in air:
\begin{eqnarray}\label{eq:recLifetimeReminder}
    \frac{1}{\tau_{v'}} &=& \frac{1}{{\tau_{0}}_{v'}} + \frac{p}{kT}\cdot \biggl[ \biggr. \nonumber \\
    &&\Bigl(0.78\cdot Q_{\textrm{N}_{2}}^{v'}(T) + 0.21\cdot Q_{\textrm{O}_{2}}^{v'}(T)\Bigr)\cdot
    \left(1-\frac{p_{\textrm{H}_{2}\textrm{O}}}{p}\right) \nonumber \\
    &&\biggl. + \frac{p_{\textrm{H}_{2}\textrm{O}}}{p}\cdot
    Q_{\textrm{H}_{2}\textrm{O}}^{v'}(T)\biggr]
\end{eqnarray}
This expression already accounts for the water vapor quenching if the water vapor partial pressure
$p_{\textrm{H}_{2}\textrm{O}}$ is given, which derives from the
relative humidity by means of the magnus
formula~\cite{Waldenmaier:2006,Murray:1967}. The temperature dependent
quenching rate constants $Q_{x}^{v'}(T)$ can be calculated
from the tabulated values in Table~\ref{tab:compQuenchRates} by
means of Eq.~(\ref{eq:QuenchRateTemp}).

\begin{figure}[b]
\includegraphics[width=\linewidth]{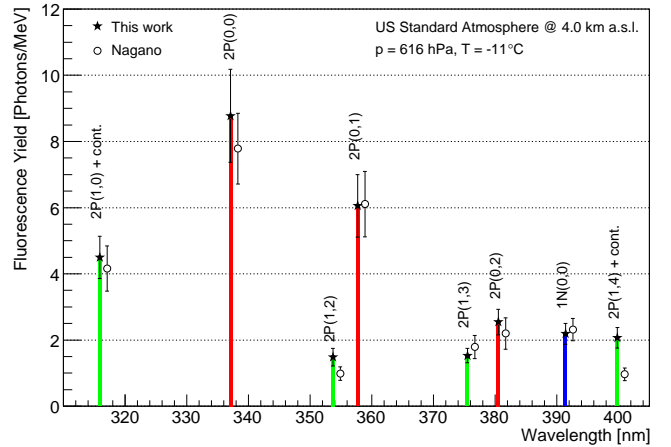}
\caption{Absolute fluorescence yield spectrum of the investigated
nitrogen transitions in dry air, at atmospheric conditions as they
appear at shower maximum, compared to the values of Nagano et al.~\cite{Nagano:2004}. The 2P(1,4) and the 2P(1,0) transitions of this work still contain contributions of neighboring bands (see text).}
\label{fig:AbsFlYieldSpectrum}
\end{figure}

On the basis of Eq.~(\ref{eq:FinalYieldReminder}) fluorescence
yield spectra can be calculated for the whole pressure and
temperature range applying to cosmic ray air shower experiments.
An example is given in Fig.~\ref{fig:AbsFlYieldSpectrum} where the
spectrum for dry air at 4~km height above sea level is compared to
the corresponding spectrum of Nagano et al.~\cite{Nagano:2004}.
The pressure and temperature at 4~km height are computed according
to the US Standard Atmosphere~\cite{USAtmo:1967} and roughly
describe the atmospheric conditions as they appear at the shower
maximum where most of the energy is deposited in the air.

At a first glance both spectra seem to agree within their errors
although there are big differences between the intrinsic yield
values in Table~\ref{tab:compIntrYields}. This is due to the
reason that the de-excitation parameters (intrinsic lifetimes, quenching rate constants) of both experiment are not exactly the same and the intrinsic yield values are calculated in the limit of zero pressure, in order to separate the excitation and de-excitation processes. In pressure ranges where the
actual measurements have been performed the agreement between both
experiments is much better since the different de-excitation parameters compensate for the differing intrinsic yield values.

The biggest difference in Fig.~\ref{fig:AbsFlYieldSpectrum} occurs for the 2P(1,4) transition, which in the present work appears to be too large. This was expected since this work's 2P(1,4) transition is still contaminated by neighboring bands and therefore is only an effective measure of all the 2P
contributions in the corresponding filter channel. Also the
2P(1,0) transition is known to be slightly contaminated by
neighboring bands, however, there seems to be no remarkable
difference compared to the spectrum of Nagano et al..

\begin{figure}[t]
\includegraphics[width=\linewidth]{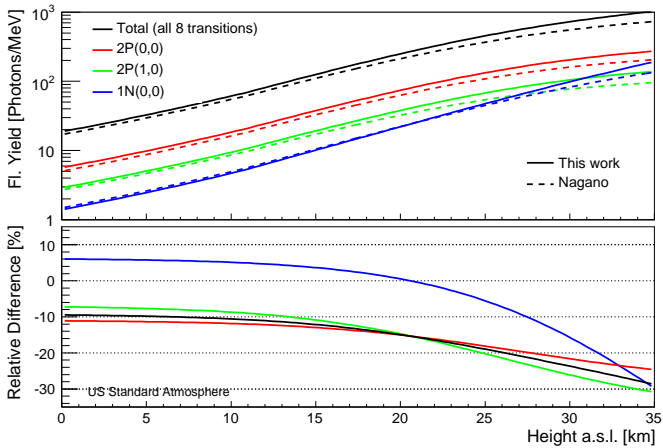}
\caption{Fluorescence yield of the three main-transitions and the
sum of all investigated transitions versus height a.s.l.. The
lower panel shows the relative differences between the values of Nagano et al. and this work in more detail.}
\label{fig:RelYieldDifferences}
\end{figure}

The differences between both spectra become more evident if
the fluorescence yield values are compared over a large
atmospheric range. In Fig.\ref{fig:RelYieldDifferences} the
fluorescence yield of the 2P(0,0), 2P(0,1) and 1N(0,0)
main-transitions as well as the sum of all 8 investigated
transitions is drawn versus the height above sea level. In the lower regions of the atmosphere this work's fluorescence yield values for the 2P transitions in average are about 10~\% larger, whereas the value of the 1N(0,0) transition appears to be 6~\% smaller than the corresponding values of Nagano et al.. These deviations are still within the systematic accuracies of both experiments. Above 10~km height the influence of the different de-excitation parameters becomes visible, leading to a slower increase of the values of Nagano et al.. Since the fluorescence light emission in EAS occurs mainly in the lower atmosphere a 10~\% difference between both fluorescence models can be expected.

\begin{figure}[t]
\includegraphics[width=\linewidth]{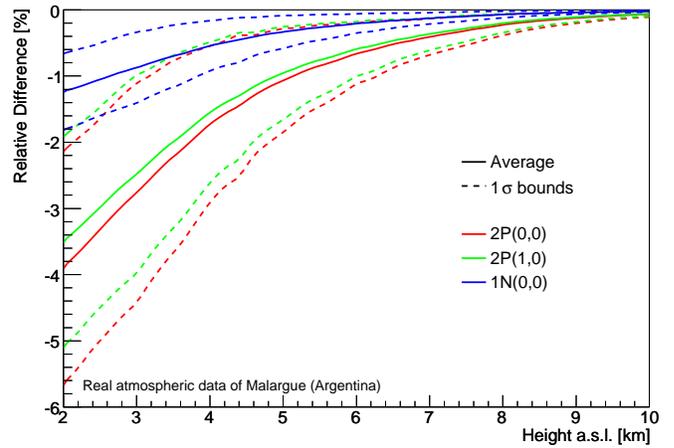}
\caption{Relative differences between this work's fluorescence
yield with and without taking into account the water vapor content
of the atmosphere. Realistic atmospheric profiles from radiosonde
launches~\cite{Keilhauer:2005,priv:Keilhauer} have been used for
this comparison.} \label{fig:yieldDiffHumidAir}
\end{figure}

Water vapor in the atmosphere leads to a reduction of the fluorescence
yield as shown in Fig.~\ref{fig:yieldDiffHumidAir}, where this work's fluorescence yield is compared for dry and humid
atmospheres. Realistic atmospheric profiles, obtained from
radiosonde measurements~\cite{Keilhauer:2005,priv:Keilhauer} at
the site of the Pierre~Auger~Observatory~\cite{Abraham:2004} in Malarg\"ue (Argentina), have been used to compute the
fluorescence yield with and without taking into account the water
vapor concentration. The solid lines correspond to the average of
187 radiosonde launches accomplished in all seasons. The dashed
lines indicate the upper and lower bounds of one standard
deviation to the average due to the varying water vapor content in
the air. At ground level the fluorescence yield of the 2P
transitions is 2~\% to 5~\% lower if the water vapor quenching is
taken into account. The 1N transition is only reduced by 1~\% to
2~\% since its quenching rate in dry air is already relatively
large.

These differences are still concealed by the systematic
uncertainties of the fluorescence yield but might become an issue
if the fluorescence yield is known with a better precision.
Currently the main source of uncertainty are the quantum- and
collection efficiencies of the PMTs which have only been estimated
according to the manufacturers specifications. Therefore, the
ongoing work is concentrating on the improvement of the absolute
calibration of the experiment using Rayleigh-scattering of a
pulsed 337~nm laser, which substitutes the electron beam in the
chamber \cite{Gonzalez:2007}. With this end-to-end calibration of
the whole setup, the absolute uncertainties of the fluorescence
yield are expected to drop below 10~\%.

\section{Acknowledgements}
The authors are very grateful to G\"unter W\"orner who constructed
the chamber and provided excellent expertise in technical matters.

\bibliographystyle{elsart-num}
\bibliography{AirLightPaper}

\begin{thebibliography}{10}
\expandafter\ifx\csname url\endcsname\relax
  \def\url#1{\texttt{#1}}\fi
\expandafter\ifx\csname urlprefix\endcsname\relax\def\urlprefix{URL }\fi

\bibitem{Abraham:2004}
J.~Abraham, et~al., {Properties and performance of the prototype instrument for
  the Pierre Auger Observatory}, Nucl. Instr. \& Meth. A 523 (2004) 50--95.

\bibitem{Springer:2005}
R.~W. Springer, et~al., {Recent results from the HiRes air fluorescence
  experiment}, Nucl. Phys. Proc. Suppl. 138 (2005) 307--309.

\bibitem{Bunner:1967}
A.~N. Bunner, Cosmic ray detection by atmospheric fluorescence, Ph.D. thesis,
  Graduate School of Cornell University (February 1967).

\bibitem{Kakimoto:1995}
F.~Kakimoto, et~al., {A Measurement of the air fluorescence yield}, Nucl.
  Instr. \& Meth. A 372 (1996) 527--533.

\bibitem{Nagano:2003}
M.~Nagano, K.~Kobayakawa, N.~Sakaki, K.~Ando, {Photon yields from nitrogen gas
  and dry air excited by electrons}, Astropart. Phys. 20 (2003) 293--309.

\bibitem{Nagano:2004}
M.~Nagano, K.~Kobayakawa, N.~Sakaki, K.~Ando, {New measurement on photon yields
  from air and the application to the energy estimation of primary cosmic
  rays}, Astropart. Phys. 22 (2004) 235--248.

\bibitem{Ave:2007}
M.~Ave, et~al., {Measurement of the pressure dependence of air fluorescence
  emission induced by electrons}, Astropart. Phys. 28 (2007) 41--57.

\bibitem{Huntemeyer:2003}
P.~Huntemeyer, An experiment to measure the air fluorescence yield in
  electromagnetic showers, AIP Conf. Proc. 698 (2004) 341--344.

\bibitem{Waldenmaier:2006}
T.~Waldenmaier, Spectral resolved measurement of the nitrogen fluorescence
  yield in air induced by electrons, Ph.D. thesis, University of Karlsruhe~(TH)
  (April 2006).
\newline\urlprefix\url{http://bibliothek.fzk.de/zb/berichte/FZKA7209.pdf}

\bibitem{priv:Ulrich}
A.~Ulrich, private communication.

\bibitem{Pearse:1976}
R.~W.~B. Pearse, A.~G. Gaydon, The Identification of Molecular Spectra, Chapman
  and Hall Ltd, 11 New Fetter Lane, London EC4P 4EE, 1976.

\bibitem{Bingel:1967}
W.~A. Bingel, Theorie der Molek{\"u}lspektren, Verlag Chemie GmbH,
  Bergsstra{\ss}e, Weinheim, 1967.

\bibitem{Haken:1992}
H.~Haken, H.~C. Wolf, Molek{\"u}lphysik und Quantenchemie, Springer-Verlag,
  Berlin Heidelberg New York, 1992.

\bibitem{Gilmore:1992}
F.~Gilmore, R.~R. Laher, P.~J. Espy, {Franck-Condon factors, r-centroids,
  electronic transition moments, and Einstein coefficients for many nitrogen
  and oxygen band systems}, J. Phys. Chem. Ref. Data 21 (1992) 1005.

\bibitem{Klepser:2004}
S.~Klepser, {Optische Elemente des AirLight-Experiments zur spektralen Messung
  der Fluoreszenzausbeute von Luft}, Master's thesis, University of
  Karlsruhe~(TH) (2004).

\bibitem{Murray:1967}
F.~W. Murray, {On the Computation of Saturation Vapor Pressure}, J. Appl.
  Meteorol. 6 (1967) 203--204.

\bibitem{Pancheshnyi:2000}
S.~V. Pancheshnyi, S.~M. Starikovskaia, A.~Y. Starikovskii, {Collisional
  dectivation of $\mathrm{N_{2}}(C^{3}\Pi_{u},v=0,1,2,3)$ states by
  $\mathrm{N_{2}}$, $\mathrm{O_{2}}$, $\mathrm{H_{2}}$ and $\mathrm{H_{2}O}$
  molecules}, Chem. Physics 262 (2000) 349--357.

\bibitem{Morozov:2005}
A.~Morozov, R.~Kr{\"u}cken, J.~Wieser, A.~Ulrich, {Gas kinitec studies using a
  table-top set-up with electron beam excitation: quenching of molecular
  nitrogen emission by water vapor}, Eur. Phys. J. D 33 (2005) 207--211.

\bibitem{Fons:1996}
T.~J. Fons, R.~S. Schappe, C.~C. Lin, {Electron-impact excitation of the second
  positive band system ($C^{3}\Pi_{u}\rightarrow B^{3}\Pi_{g}$) and the
  $C^{3}\Pi_{u}$ electronic state of the nitrogen molecule}, Phys. Rev. A
  53~(4) (1996) 2239--2247.

\bibitem{Agostinelli:2002}
S.~Agostinelli, et~al., {GEANT4: A simulation toolkit}, Nucl. Instr. \& Meth. A
  506 (2003) 250--303.

\bibitem{Seltzer:1982}
S.~M. Seltzer, B.~M. J., {Evaluation of the Collision Stopping Power of
  Elements and Compounds for Electrons and Positrons}, Int. J. Appl. Radiat.
  Isot. 33 (1982) 1189--1218.

\bibitem{USAtmo:1967}
{U.S. Standard Atmosphere}, U.S. Government Printing Office, Washington, D.C.,
  1976.

\bibitem{Keilhauer:2005}
B.~Keilhauer, et~al., Atmospheric profiles at the southern pierre auger
  observatory and their relevance to air shower measurement, Proc. 29th Int.
  Cosmic Ray Conf., Pune 7 (2005) 123.

\bibitem{priv:Keilhauer}
B.~G. Keilhauer, private communication.

\bibitem{Gonzalez:2007}
D.~M. Gonzalez, et~al., {Laser Calibration of the Air Fluorescence Yield
  Experiment AIRLIGHT}, Proc. 30th Int. Cosmic Ray Conf., Merida.

\end{thebibliography}

\end{document}